\documentclass[12pt]{article}
\usepackage{amssymb,amsmath,bm}
\usepackage{epsf}
\usepackage{epsfig}
\def\dfrac#1#2{{\displaystyle {#1 \over #2}}}

\def\simge{\mathrel{\rlap{\raise 0.511ex \hbox{$>$}}{\lower 0.511ex \hbox{$\sim$}}}}
\def\simle{\mathrel{\rlap{\raise 0.511ex \hbox{$<$}}{\lower 0.511ex \hbox{$\sim$}}}}
\def\slash#1{\setbox0=\hbox{$#1$}\dimen0=\wd0
      \setbox1=\hbox{/} \dimen1=\wd1 \ifdim\dimen0>\dimen1
      \rlap{\hbox to \dimen0{\hfil/\hfil}} #1                        \else
      \rlap{\hbox to \dimen1{\hfil$#1$\hfil}}
      /   \fi}

\newcommand{\newsection}[1]{\section{#1}\setcounter{equation}{0}}

\newcommand{\lsim}{
\mathrel{\hbox{\rlap{\hbox{\lower4pt\hbox{$\sim$}}}\hbox{$<$}}}}

\newcommand{\gsim}{
\mathrel{\hbox{\rlap{\hbox{\lower4pt\hbox{$\sim$}}}\hbox{$>$}}}}

\newcommand{\vcb}{|V_{cb}|}

\def\eps{\varepsilon}

\newcommand{\tev}{\, {\rm TeV}}
\newcommand{\gev}{\, {\rm GeV}}
\newcommand{\mev}{\, {\rm MeV}}

\newcommand{\GF}{\dfrac{G_{ F}}{\sqrt 2}}
\newcommand{\Heff}{{\cal H}_\text{ eff}}

\newcommand{\mtb}{\overline{m}_{\rm t}}
\newcommand{\mcb}{\overline{m}_{\rm c}}

\newcommand{\mw}{M_{\rm W}}

\allowdisplaybreaks[2]

\newcommand{\be}{\begin{equation}}
\newcommand{\ee}{\end{equation}}
\newcommand{\bea}{\begin{eqnarray}}
\newcommand{\eea}{\end{eqnarray}}
\newcommand{\nn}{\nonumber}
\newcommand{\bi}{\begin{itemize}}
\newcommand{\ei}{\end{itemize}}
\newcommand{\ord}{{\cal O}}

\begin{document}
\begin{titlepage}
\vspace*{-0.5truecm}

\begin{flushright}
TUM-HEP-630/06\\
\end{flushright}

\vspace{1truecm}

\begin{center}
\boldmath

{\Large\textbf{Particle-Antiparticle Mixing, $\varepsilon_K$, $\Delta
    \Gamma_q$, $A_{\rm SL}^q$,\vspace{0.3truecm}
    $A_{\rm CP}(B_d \rightarrow \psi K_S)$, $A_{\rm CP}(B_s\vspace{0.4truecm}
    \rightarrow \psi \phi)$ and $B\rightarrow X_{s,d}\gamma$ in the
    Littlest Higgs Model with T-Parity}}

\unboldmath
\end{center}

\vspace{0.4truecm}

\begin{center}
{\large\bf Monika Blanke, Andrzej J.~Buras, Anton Poschenrieder,\\
 Cecilia  Tarantino, Selma Uhlig and Andreas Weiler
}
\vspace{0.4truecm}

 {\sl Physik Department, Technische Universit\"at M\"unchen,
D-85748 Garching, Germany}

\end{center}

\vspace{0.6cm}
\begin{abstract}
\vspace{0.2cm}\noindent
We calculate a number of observables related to particle-antiparticle 
mixing in 
the Littlest Higgs model with T-parity (LHT). The resulting effective 
Hamiltonian for $\Delta F=2$ transitions agrees with the one of Hubisz 
et al., but our phenomenological analysis goes far beyond the one of these 
authors. In particular, we point out that the presence of mirror fermions 
with new flavour and CP-violating interactions allows to remove the 
possible Standard Model (SM) discrepancy between the CP asymmetry $S_{\psi K_S}$ and large 
values of $|V_{ub}|$ and to obtain for the mass difference 
$\Delta M_s< (\Delta M_s)_{\rm SM}$ as suggested by the recent result by
 the CDF collaboration. 
We also identify a scenario  in which 
simultaneously significant enhancements of the CP asymmetries 
$S_{\psi \phi}$ and $A_{\rm SL}^q$ relative to the SM
are possible, while satisfying all existing 
constraints, in particular from the $B\to X_s\gamma$ decay and 
$A_{\rm CP}(B\to X_s\gamma)$ that are presented in the LHT model here
for the first time. In another scenario the second, non-SM, value for the 
angle $\gamma=-(109\pm16)^\circ$ from tree level decays, although unlikely,
can be made consistent with all existing data with the help of mirror fermions.
We present a number of correlations between the observables in question
and study the
implications of our results for the mass spectrum and the weak mixing 
matrix of mirror fermions. In the most interesting scenarios, the latter
one turns out to have a hierarchical structure that differs
significantly from 
the CKM one.

\end{abstract}

\end{titlepage}

\setcounter{page}{1}
\pagenumbering{arabic}

\newsection{Introduction}
\label{sec:intro}

One of the most important messages that will be hopefully provided
in the coming years by LHC and later by ILC is the detailed
information about the electroweak symmetry breaking (EWSB) and the
origin of the hierarchy of quark masses and their hierarchical
flavour and CP-violating interactions. While
supersymmetry~\cite{susy} appears at present to be the leading
candidate for a self-consistent incorporation of the Higgs
mechanism into the framework of gauge theories, recent proposals
like Little Higgs models~\cite{oldLH,LHreview}, Extra dimension
models~\cite{ADD,RS}, gauge-Higgs unification
models~\cite{gaugehiggsgeneral,sss} 
and
improved versions of technicolour~\cite{TC,TCreview} and top
colour~\cite{topc} have still potential to provide at least
partial solutions to EWSB and to shed light on the hierarchical
structure of flavour violating interactions. Each of these
proposals introduces new particles below $1 \tev$
 or slightly above it with often significant impact of
their contributions on electroweak precision studies and
FCNC processes.

Among the most popular non-supersymmetric models in question are
the Little Higgs models of which the so-called Littlest Higgs
model \cite{l2h} has been studied most extensively in the
literature (see \cite{LHreview} and references therein). In this model in addition to the
Standard Model (SM) particles, new charged heavy vector bosons
($W_H^\pm$), a neutral heavy vector boson ($Z_H^0$), a heavy
photon ($A_H$), a heavy top quark ($T_+$) and a triplet of scalar
heavy particles ($\Phi$) are present.

In the original Littlest Higgs model (LH), the custodial $SU(2)$
symmetry, of fundamental importance for electroweak precision studies, is
unfortunately broken already at tree level, implying that the
relevant scale of new physics, $f$, must be at least $2-3 \tev$ in
order to be consistent with electroweak precision data \cite{Logan}-\cite{PHEN6}.
As a
consequence the contributions of the new particles to FCNC
processes turn out to be at most $10-20\%$~\cite{BPU}-\cite{bsgl2h},
which will not be easy to distinguish from the SM in view of
experimental and theoretical uncertainties.

More promising and more interesting from the point of view
of FCNC processes is the Littlest Higgs model with a
discrete symmetry (T-parity)~\cite{tparity} under which all new particles
listed above, except $T_+$, are odd and do not contribute to
processes with external SM quarks (even under T-parity) at tree
level. As a consequence the new physics scale $f$ can be
lowered down to $1 \tev$ and even below it, without violating
electroweak precision constraints~\cite{mH}.

A consistent and phenomenologically viable Littlest Higgs model
with T-parity (LHT) requires the introduction of three doublets of
``mirror quarks'' and three doublets of ``mirror leptons'' which
are odd under T-parity, transform vectorially under $SU(2)_L$ and
can be given a large Dirac mass. Moreover, there is an additional
heavy $T_-$ quark that is also odd under T-parity
\cite{mirror}.\footnote{In \cite{0510225}, an alternative way of
  implementing T-parity in the top sector has been proposed, where $T_+$ is absent.}

In the first phenomenological studies of the LHT
model~\cite{Hubisz:2004ft} the contributions of mirror fermions to
physical observables have not been considered,
while their impact on electroweak precision tests has been investigated in~\cite{mH}.
More recently, in an
interesting paper by Hubisz et al.~\cite{Hubisz} the role of
mirror fermions in neutral meson mixing in the $K$, $B$ and $D$
systems has been studied in some detail. The main messages
of~\cite{Hubisz} are:

\begin{itemize}
\item There are new flavour violating interactions in the
       mirror quark sector, which can be parameterized by two CKM-like mixing
       matrices $V_{Hd}$ and $V_{Hu}$, relevant for the processes with
       external light down-type quarks and up-type quarks, respectively.
       These two matrices are related through $V_{Hu}^\dagger V_{Hd} =
       V_\text{CKM}$. Similar comments
       apply to the mirror lepton sector.
\item  The spectrum of mirror quarks must be generally quasi-degenerate, if
   $\ord(1)$ mixing angles are allowed in the new mixing matrices, but
there exist regions of parameter space, where
only a loose degeneracy is necessary in order to satisfy
constraints coming from particle-antiparticle mixing.
\end{itemize}

The recent measurements of the $B^0_s-\bar B^0_s$ mass
difference $\Delta M_s$ by the CDF and D{\O} collaborations
\cite{CDFnew,D0}, that turns out to be close to the SM value, puts clearly
an additional constraint on the model in question.

In the present paper we confirm the analytic expressions for
the effective Hamiltonians for $K^0-\bar K^0$, $B_d^0-\bar B_d^0$ and $B_s^0-\bar B_s^0$ mixings
presented in~\cite{Hubisz} and we generalize the analysis of these
authors to other quantities that allow a deeper insight into
the flavour structure of the LHT model. However, before
listing the new aspects of our paper relatively to~\cite{Hubisz}, let us
emphasize a few points about the model in question that have
not been stated so far in the literature.

While the original LH model belongs to the class of models
with Minimal Flavour Violation (MFV)~\cite{UUT}-\cite{mfvearly}, 
this is certainly not the case of the LHT model
where the presence of the matrices $V_{Hd}$ and $V_{Hu}$ in the mirror
quark sector introduces new flavour and CP-violating
interactions that could have a very different pattern from the
ones present in the SM.

One should also emphasize that the manner in which the LHT model
goes beyond the MFV scenario differs from the frameworks
studied in~\cite{BFRS} and~\cite{NMFV}, where the modification of the flavour
structure is connected dominantly to the third generation
of quarks. Here, the new flavour violating interactions come
simply from another sector that couples weakly to ordinary
fermions and in principle all generations of mirror fermions
can contribute to FCNC processes with equal strength.

The beauty of this model, when compared with other models with
non-minimal flavour violating interactions, like general
MSSM, is a relatively small number of new parameters and the
fact that the local operators involved are the same as in
the SM. Therefore the non-perturbative uncertainties,
present in certain quantities already in the SM, are the
same in the LHT model. Consequently the departures from the
SM are entirely due to short distance physics that can be
calculated within perturbation theory. In stating this we
are aware of the fact that we deal here with an effective
field theory whose ultraviolet completion has not been
specified, with the consequence that at a certain level of
accuracy one has to worry about the effects coming from the
cut-off scale $\Lambda\sim 4\pi f$. We will assume that in the case of
particle-antiparticle mixing and $B\to X_s\gamma$ such effects are small.

So what is new in our paper relatively to~\cite{Hubisz}?

\begin{itemize}
\item   While the authors of~\cite{Hubisz} analyzed only the mass
       differences $\Delta M_K$, $\Delta M_d$, $\Delta M_s$, $\Delta
       M_D$ and the CP violating parameter $\varepsilon_K$, we include
       in our analysis also the CP asymmetries $A_{\rm CP}(B_d
       \rightarrow \psi K_S)$, $A_{\rm CP}(B_s \rightarrow \psi \phi)$
       and $A_{\rm SL}^q$, and the width difference $\Delta\Gamma_q$,
       that are theoretically cleaner than the
       quantities considered in~\cite{Hubisz}.
\item  Equally important, we present for the first time the
expressions for the $B\rightarrow X_{s,d}\gamma$ decay within the LHT model.
As $B\rightarrow X_s\gamma$ has played already an important role in
constraining other extensions of the SM and is experimentally measured with
good accuracy, it is mandatory to study it in the LHT model as well.
In this context we also calculate the corresponding CP asymmetries.
\item Our analysis of the mixing induced CP asymmetries $A_{\rm CP}(B_d \rightarrow \psi K_S)$ and $A_{\rm CP}(B_s\linebreak \rightarrow \psi \phi)$
illustrates very clearly that with mirror fermions at work
these asymmetries do \emph{not} measure the phases $-\beta$ and $-\beta_s$ of the
CKM elements $V_{td}$ and $V_{ts}$, respectively.
\item  This has two interesting consequences: first, the
possible ``discrepancy'' between the values of $\sin 2\beta $ following
directly from $A_{\rm CP}(B_d \rightarrow \psi K_S)$ and indirectly from the usual analysis of
the unitarity triangle involving $\Delta M_q$, $\varepsilon_K$ and $|V_{ub}/V_{cb}|$ can be
avoided within the LHT model. Second, the asymmetries
 $A_{\rm CP}(B_s \rightarrow \psi \phi)$ and $A_{\rm SL}^q$ can
be significantly enhanced over the SM expectations.
\item In connection with the recent measurement of $\Delta M_s$ by the CDF
collaboration \cite{CDFnew}, that although close to the SM value,
is somewhat lower than expected, we investigate for which set of
parameters of the LHT model $\Delta M_s$ can be lower than
$(\Delta M_s)_{\rm SM}$ while simultaneously solving the ``$\sin
2\beta$'' problem mentioned above.
\item   We also find that the usual relation between $\Delta M_d/\Delta M_s$ and $|V_{td}/V_{ts}|$
characteristic for all models with MFV is no longer
satisfied.
\item We introduce the concept of the ``mirror unitarity
triangle'' which is also useful when the analysis is 
generalized to include rare $K$ and $B$ decays \cite{BBPRTUW}.
\item We also investigate whether the second, non-MFV, solution for
  $\gamma=-109^\circ$ from tree level decays can be made consistent
  with all available data. 
\item  Finally, we present explicit formulae for the
contributions of the T-even sector, that in the model in
question are entirely dominated by the contributions of the
heavy $T_+$ quark. We emphasize that these contributions cannot
be neglected for values of the parameter $x_L> 0.5$ and in the limit of
exactly degenerate mirror fermions constitute the only new
contributions in this model.
\end{itemize}

Our paper is organized as follows. In Section~\ref{sec:model} we
summarize those ingredients of the LHT model that are of relevance
for our analysis and we introduce mirror unitarity triangles.
Section~\ref{sec:mix} is devoted to the particle-antiparticle
mixings, $\varepsilon_K$, the asymmetries $A_{\rm CP}(B_d
\rightarrow \psi K_S)$, $A_{\rm CP}(B_s \rightarrow \psi \phi)$,
$A_{\rm SL}^q$, the width differences $\Delta \Gamma_q$ and in
particular to the ratio $\Delta M_d/\Delta M_s$. We collect in
this section a number of formulae that should be useful also for other
models. In Section~\ref{sec:bsg} we calculate the branching ratios
for $B\rightarrow X_s\gamma$ and $B\rightarrow X_d\gamma$ and the 
 corresponding CP asymmetries. In
Section \ref{sec:goals} we outline our strategy and our goals for
the numerical analysis of Section
\ref{sec:beyond}.
 In Section~\ref{sec:MFV} we discuss the benchmark scenarios for 
the parameters of 
the LHT model, which we explore in
Section~\ref{sec:beyond}, where the correlations between various
observables can be studied more explicitly than it is possible in the recent model
independent analyses in~\cite{NMFV,BBGT}-\cite{Datta}. 
 It is
in this section where we address the possible discrepancy
between the indirect and direct determinations of the angle
$\beta$ in the UT, its resolution within the LHT model, the
enhancements of $A_{\rm CP}(B_s \rightarrow \psi \phi)$ and 
$A_{\rm SL}^q$  and the
size of the corrections to the MFV result for $\Delta M_d/\Delta
M_s$. A highlight of this section is also the analysis of the
mirror fermion contributions to $\Delta M_s$ in view of the recent
measurements of $B^0_s - \bar B^0_s$ mixing~\cite{CDFnew,D0} and
its possible correlation with ${Br}(B\rightarrow X_s\gamma)$. Also the
rescue of the non-SM solution for $\gamma$ with the help of mirror
fermions is demonstrated in this section. In
Section~\ref{sec:DD} we discuss briefly the $D^0-\bar D^0$
mixing. 
Finally, in
Section~\ref{sec:summary} we conclude our paper with a list of
messages resulting from our analysis and with a brief outlook. Few
technical details are relegated to the Appendices.

\newsection{General Structure of the LHT Model}
\label{sec:model}

A detailed description of the LHT model can be found e.\,g. in
\cite{Hubisz:2004ft}. Here we just want to state briefly the
ingredients needed for our analysis.

\subsection{Gauge Boson Sector}
\label{subsec:2.1}
\subsubsection{T-even Sector}
\label{subsubsec:2.1.1}

The T-even electroweak gauge boson sector \cite{l2h}  consists only of SM
electroweak gauge bosons
\begin{equation}\label{2.1}
W_L^\pm\,,\qquad Z_L\,,\qquad A_L\,,
\end{equation}
with masses given to lowest order in $v/f$ by
\begin{equation}\label{2.2}
M_{W_L}=\frac{gv}{2}\,,\qquad
M_{Z_L}=\frac{M_{W_L}}{\cos{\theta_W}}\,,\qquad M_{A_L}=0\,,
\end{equation}
where $\theta_W$ is the weak mixing angle. T-parity ensures that
the second relation in \eqref{2.2} is satisfied at tree level to
all orders in $v/f$. Only $W_L^\pm$ will be present in the
discussion of $\Delta F=2$ processes while both $A_L$ and
$W_L^\pm$ enter the $B\rightarrow X_s\gamma$ decay.

\subsubsection{T-odd Sector}
\label{subsubsec:2.1.2}
The T-odd gauge boson sector~\cite{l2h} consists of three heavy
``partners'' of the SM gauge bosons in \eqref{2.1}:
\begin{equation}\label{2.3}
W_H^\pm\,,\qquad Z_H\,,\qquad A_H\,,
\end{equation}
with masses given to lowest order in $v/f$ by
\begin{equation}\label{2.4}
M_{W_H}=gf\,,\qquad M_{Z_H}=gf\,,\qquad
M_{A_H}=\frac{g'f}{\sqrt{5}}\,.
\end{equation}
All three gauge bosons will be present in our analysis. Note
that
\begin{equation}\label{2.4a}
M_{A_H}=\frac{\tan{\theta_W}}{\sqrt{5}}M_{W_H}\simeq\frac{M_{W_H}}{4.1}\,.
\end{equation}

\subsection{Fermion Sector}
\label{subsec:2.2}
\subsubsection{T-even Sector}
\label{subsubsec:2.2.1} The T-even fermion sector~\cite{l2h}
consists of the SM quarks and leptons and a colour triplet heavy
quark $T_+$ that is, to leading order in $v/f$, singlet under
$SU(2)_L$ and has the mass
\begin{equation}\label{2.5}
m_{T_+}=\frac{f}{v}\frac{m_t}{\sqrt{x_L(1-x_L)}}\,,\quad x_L =\frac{\lambda_1^2}{\lambda_1^2+\lambda_2^2}\,.
\end{equation}
Here $\lambda_1$ is the Yukawa coupling in the $(t, T_+)$ sector and
$\lambda_2$ parameterizes the mass term of $T_+$.

\subsubsection{T-odd Sector}
\label{subsubsec:2.2.2}
The T-odd fermion sector~\cite{mirror} consists first of all of three
generations of mirror quarks and leptons with vectorial
couplings under $SU(2)_L$. In this paper only mirror quarks are
relevant. We will denote them by 
  \begin{equation}\label{2.6}
\begin{pmatrix} u^1_{H}\\d^1_{H} \end{pmatrix}\,,\qquad
\begin{pmatrix} u^2_{H}\\d^2_{H} \end{pmatrix}\,,\qquad
\begin{pmatrix} u^3_{H}\\d^3_{H} \end{pmatrix}\,,
\end{equation}
with their masses satisfying to first order in $v/f$
\begin{equation}\label{2.7}
m^u_{H1}=m^d_{H1}\,,\qquad m^u_{H2}=m^d_{H2}\,,\qquad
m^u_{H3}=m^d_{H3}\,.
\end{equation}

The T-odd fermion sector contains also a T-odd heavy quark $T_-$,
which will not enter our analysis for reasons given in Appendix
\ref{sec:appA}. For completeness, we quote the expression for its mass,
\begin{equation}\label{2.8}
m_{T_-}=\lambda_2 f =\frac{f}{v}\frac{m_t}{\sqrt{x_L}}\,.
\end{equation}
In principle a lower bound on $m_{T_-}$ like $m_{T_-}>500\,\gev$ could
set an upper bound on $x_L$ for fixed
$f$, but it turns out that the electroweak precision constraints are
more important \cite{mH}.

\subsection{Scalar Triplet}
\label{subsec:2.3} For completeness we mention that also a Higgs
triplet $\Phi$ belongs to the T-odd sector. The charged Higgs
$\phi^\pm$, as well as the neutral Higgses $\phi^0,\;\phi^P$, are
relevant in principle for the decays considered here, but their
effects turn out to be of higher order in $v/f$ as explained in Appendix
\ref{sec:appA}. Their mass is given by
\begin{equation}\label{2.9}
m_{\Phi}=\sqrt{2}m_H\frac{f}{v}\,,
\end{equation}
where $m_H$ is the mass of the SM Higgs. As pointed out in~\cite{mH},
$m_H$ in the LHT model can be significantly larger than in
supersymmetry.

\subsection{Weak Mixing in the Mirror Sector}
\label{subsec:2.4}
As discussed in detail in~\cite{Hubisz}, one of the important
ingredients of the mirror sector is the existence of four CKM-like unitary
mixing matrices, two for mirror quarks and two
for mirror leptons:
\begin{equation}\label{2.10}
V_{Hu}\,,\quad V_{Hd}\,,\qquad V_{H\ell}\,,\quad V_{H\nu}\,.
\end{equation}
They satisfy
\begin{equation}\label{2.11}
V_{Hu}^\dagger V_{Hd}=V_\text{CKM}\,,\qquad V_{H\nu}^\dagger
V_{H\ell}=V_\text{PMNS}^\dagger\,,
\end{equation}
where in $V_\text{PMNS}$~\cite{pmns} the Majorana phases are set
to zero as no Majorana mass term has
been introduced for the right-handed neutrinos. The mirror mixing
matrices in \eqref{2.10} parameterize flavour violating
interactions between SM fermions and mirror fermions
that are mediated by the heavy gauge bosons $W_H$, $Z_H$ and $A_H$.
The notation in \eqref{2.10} indicates which of the light fermions
of a given electric charge participates in the interaction.

Thus $V_{Hd}$, the most important mixing matrix in the present
paper, parameterizes the interactions of light $d^j$-quarks with
heavy mirrors $u^i_{H}$ that are mediated by $W_H$. It also
parameterizes the flavour interactions between $d^j$ and $d^i_{H}$
mediated by $Z_H$ and $A_H$. Feynman rules for these interactions
can be found in \cite{Hubisz}. We have confirmed those which we needed 
for the present paper. $V_{Hu}$, relevant for $D^0-\bar
D^0$ mixing, parameterizes on the other hand the interactions of
the light $u$-type quarks with the mirror fermions. Similar
comments apply to $V_{H\nu}$ and $V_{H\ell}$.

In the course of our analysis of $\Delta S=2$ and $\Delta B=2$ processes and in
the case of $B\rightarrow X_s\gamma$ it will be useful to introduce the following
quantities ($i=1,2,3$):
\begin{equation}\label{2.12}
\xi_i=V^{*is}_{Hd}V^{id}_{Hd}\,,\qquad
\xi_i^{(d)}=V^{*ib}_{Hd}V^{id}_{Hd}\,,\qquad
\xi_i^{(s)}=V^{*ib}_{Hd}V^{is}_{Hd}\,,
\end{equation}
that govern $K^0-\bar K^0$, $B_d^0-\bar B_d^0$ and $B_s^0-\bar B_s^0$ mixings, respectively. $\xi_i^{(s)}$ are also
relevant for $B\rightarrow X_s\gamma$.

In~\cite{Hubisz} and consequently in the first version of this paper, 
$V_{Hd}$ was parameterized in the
same way  as the CKM matrix~\cite{ckm}, in terms of three angles
$\theta_{12}^d$, $\theta_{23}^d$, $\theta_{13}^d$ and
one phase $\delta^d_{13}$.
In \cite{SHORT}, it was pointed out for the first time that $V_{Hd}$ contains
not only one but three phases.
In short, the reason for the appearance of two additional phases relative to
the CKM matrix is as follows.
 $V_\text{CKM}$ and $V_{Hd}$ are both unitary
matrices containing three real angles and six complex phases.
Varying independently the phases of ordinary up- and down-quark states
allows us to rotate five phases away from $V_\text{CKM}$ (an over-all phase
change of all the quark states leaves $V_\text{CKM}$ invariant).
In rotating phases away from $V_{Hd}$, then, one can still act on only three
mirror states, thus obtaining for $V_{Hd}$ a parameterization in terms of
three mixing angles and three phases.

Following~\cite{SHORT} we will parameterize $V_{Hd}$ generalizing
the usual CKM parameterization, as a product of three rotations, and
introducing a complex phase in any of them, thus obtaining
\be
\addtolength{\arraycolsep}{3pt}
V_{Hd}= \begin{pmatrix}
1 & 0 & 0\\
0 & c_{23}^d & s_{23}^d e^{- i\delta^d_{23}}\\
0 & -s_{23}^d e^{i\delta^d_{23}} & c_{23}^d\\
\end{pmatrix}\,\cdot
 \begin{pmatrix}
c_{13}^d & 0 & s_{13}^d e^{- i\delta^d_{13}}\\
0 & 1 & 0\\
-s_{13}^d e^{ i\delta^d_{13}} & 0 & c_{13}^d\\
\end{pmatrix}\,\cdot
 \begin{pmatrix}
c_{12}^d & s_{12}^d e^{- i\delta^d_{12}} & 0\\
-s_{12}^d e^{i\delta^d_{12}} & c_{12}^d & 0\\
0 & 0 & 1\\
\end{pmatrix}\ee
Performing the product one obtains the expression
\be\label{2.12a}
\addtolength{\arraycolsep}{3pt}
V_{Hd}= \begin{pmatrix}
c_{12}^d c_{13}^d & s_{12}^d c_{13}^d e^{-i\delta^d_{12}}& s_{13}^d e^{-i\delta^d_{13}}\\
-s_{12}^d c_{23}^d e^{i\delta^d_{12}}-c_{12}^d s_{23}^ds_{13}^d e^{i(\delta^d_{13}-\delta^d_{23})} &
c_{12}^d c_{23}^d-s_{12}^d s_{23}^ds_{13}^d e^{i(\delta^d_{13}-\delta^d_{12}-\delta^d_{23})} &
s_{23}^dc_{13}^d e^{-i\delta^d_{23}}\\
s_{12}^d s_{23}^d e^{i(\delta^d_{12}+\delta^d_{23})}-c_{12}^d c_{23}^ds_{13}^d e^{i\delta^d_{13}} &
-c_{12}^d s_{23}^d e^{i\delta^d_{23}}-s_{12}^d c_{23}^d s_{13}^d
e^{i(\delta^d_{13}-\delta^d_{12})} & c_{23}^d c_{13}^d\\
\end{pmatrix}
\ee
As in the case of the CKM matrix 
the angles $\theta_{ij}^d$ can all be made to lie in the first quadrant 
with $0\le \delta^d_{12}, \delta^d_{23}, \delta^d_{13},\le 2\pi$.
The matrix $V_{Hu}$ is then determined through 
$V_{Hu}=V_{Hd}V_\text{CKM}^\dagger$. 

The matrix $V_{Hd}$ depends on
six parameters that have to be determined in flavour violating
processes. In Section~\ref{sec:beyond} we will outline a strategy
for this determination. As in the case of the determination of the
parameters of the CKM matrix, also here unitarity triangles could
play in the future a useful role. On the other hand the structure 
of the matrix $V_{Hd}$ can differ in principle by much from the 
structure of the CKM matrix and using approximations like the 
Wolfenstein parameterization should be avoided in order to satisfy 
unitarity exactly.

\subsection{Mirror Unitarity Triangles}
\label{subsec:2.5}
The unitarity of the $V_{Hd}$ matrix allows to construct six
unitarity triangles. The three most important correspond to the unitarity relations
\begin{gather}
\xi_1+\xi_2+\xi_3=0\qquad(K^0-\bar K^0)\label{2.13}\,,\\
\xi_1^{(d)}+\xi_2^{(d)}+\xi_3^{(d)}=0\qquad(B_d^0-\bar B_d^0)\label{2.14}\,,\\
\xi_1^{(s)}+\xi_2^{(s)}+\xi_3^{(s)}=0\qquad(B_s^0-\bar B_s^0)\label{2.15}\,.
\end{gather}

In the SM, the hierarchical structure of the elements of the
CKM matrix implies rather squashed unitarity triangles in
the $K^0-\bar K^0$ and $B_s^0-\bar B_s^0$ systems with the famous unitarity triangle in
the $B_d^0-\bar B_d^0$ system, corresponding to
\be\label{SMUT}
V_{ud}V^*_{ub}+V_{cd}V^*_{cb}+V_{td}V^*_{tb}=0\,,
\ee
having all sides of the same order of magnitude.

We have clearly no idea at present what the shapes of the mirror
unitarity triangles are. The lessons from neutrino physics teach
us that they could be very different from the ones encountered in
the SM. In Section~\ref{sec:beyond} we will see that in the most interesting 
scenarios the structure of $V_{Hd}$ is very different from the CKM one 
implying significantly different mirror unitarity triangles than the one 
following from (\ref{SMUT}). This issue is also discussed in
\cite{BBPRTUW}, where our analysis is generalized to rare $K$ and
$B$ decays.

\subsection{The Parameters of the LHT Model}
\label{subsec:2.6}
The new parameters in the LHT model, relevant for the present study, are
\begin{equation}\label{2.16}
f\,,\quad x_L\,,\quad m_{H1}\,,\quad m_{H2}\,,\quad m_{H3}\,,\quad \theta_{12}^d\,,\quad \theta_{13}^d\,,\quad \theta_{23}^d\,,\quad \delta_{12}^d\,\quad \delta_{13}^d\,\quad \delta_{23}^d\,.
\end{equation}

The determination of all these parameters with the help of
flavour violating processes is clearly a formidable task. On
the other hand once LHC starts its operation and the new
particles present in the LHT model are discovered, we
will determine $f$ from $M_{W_H}$, $M_{Z_H}$ or $M_{A_H}$ and $x_L$
from $m_{T_-}$ or $m_{T_+}$. Similarly $m_{Hi}$ will be measured.

Since the CKM parameters can be determined independently of the
LHT contributions from tree level decays during the LHC era,
the only remaining free parameters among the ones listed in
\eqref{2.16} are $\theta_{ij}^d$ and $\delta_{ij}^d$. They can be,
similarly to the parameters of the CKM matrix, determined in
flavour violating processes. In this manner also mirror
unitarity triangles will be constructed.

However, in contrast to the CKM parameters, the six
parameters of the $V_{Hd}$ matrix cannot be determined with the
help of tree level decays. In fact tree level decays are of
no help here because T-parity forbids the contributions
of mirror fermions to these decays at tree level. This
is a welcome feature for the determination of the CKM
parameters from tree level decays independently of the
presence of mirror fermions and T-odd particles, but the
determination of the parameters of $V_{Hd}$ can only
be done with the help of loop induced decays, unless decays
of mirror fermions to light fermions can be measured one
day.

In Section~\ref{sec:beyond} we will indicate how the
determination of the matrix $V_{Hd}$ could be done with
the help of the processes considered in the present paper.
Generalizations to include rare $K$ and $B$ decays in this
determination has been very recently presented in \cite{BBPRTUW}.

Clearly as the first five parameters in \eqref{2.16} are not known
at present, we will only be able to study correlations
between all these parameters that are implied by the available data.

\newsection{Particle-Antiparticle Mixing and CP Violation}
\label{sec:mix}
\subsection{T-even Sector}
\label{subsec:3.1}

The contribution of the T-even sector can be directly extracted 
from~\cite{BPU}.
Including the SM box diagrams the
effective Hamiltonian for $\Delta S=2$ transitions can be written
as follows~\cite{BPU}:
\begin{equation}\label{3.1}
\left[\Heff(\Delta S=2)\right]_\text{even}=
\frac{G_F^2}{16\pi^2}M_{W_L}^2\left[\lambda_c^2\eta_1S_c+\lambda_t^2\eta_2S_t+2\lambda_c\lambda_t\eta_3S_{ct}\right](\bar s d)_{V-A}(\bar s d)_{V-A}\,,
\end{equation}
where $\lambda_i=V_{is}^*V_{id}$. In the case of $B_d^0-\bar B_d^0$ and $B_s^0-\bar B_s^0$ mixing the first and the
last term can be neglected and one finds ($q=d,s$)
\begin{equation}\label{3.2}
\left[\Heff(\Delta B=2)\right]_\text{even}=
\frac{G_F^2}{16\pi^2}M_{W_L}^2\lambda_t^{(q)2}\eta_BS_t(\bar bq)_{V-A}(\bar bq)_{V-A}\,,
\end{equation}
where $\lambda_t^{(q)}=V_{tb}^*V_{tq}$. The factors $\eta_i$ are QCD corrections to which we will return
in Section~\ref{subsec:3.5}.

Writing
\begin{equation}\label{3.4}
S_t=S_0(x_t)+\Delta S_t+\Delta S_{TT},\quad S_c=S_0(x_c)+\Delta S_c\,,\quad
S_{ct}=S_0(x_c,x_t)+\Delta S_{ct}\,,
\end{equation}
with $S_0$ being the SM contributions, we find directly from~\cite{BPU}
\begin{equation}\label{3.6}
\Delta S_t=-2\frac{v^2}{f^2}x_L^2P_1(x_t,x_T)\,,\quad \Delta S_c=0\,,\quad
\Delta S_{ct}=-\frac{v^2}{f^2}x_L^2P_2(x_c,x_t,x_T)\,,
\end{equation}
\be\label{TT}
\Delta S_{TT}\simeq \frac{v^2}{f^2}\frac{x_L^3}{1-x_L}\frac{x_t}{4},
\ee
with $P_1(x_t,x_T)$ and $P_2(x_c,x_t,x_T)$ calculated in~\cite{BPU} and given for completeness
in Appendix B. Here,
\be
x_c=\frac{m_c^2}{M_{W_L}^2}\,,\qquad x_t=\frac{m_t^2}{M_{W_L}^2}\,,\qquad x_T=\frac{m_{T_+}^2}{M_{W_L}^2}\,.
\label{eq:xtxT}
\ee

The contribution of the T-even sector to the off-diagonal
element $M^K_{12}$ in the neutral $K$-meson mass matrix is then given
as follows
\begin{equation}\label{3.7}
\left(M_{12}^K\right)_\text{even}=\frac{G_F^2}{12\pi^2}F_K^2\hat
B_K m_K M_{W_L}^2\left(\overline{M_{12}^K}\right)_\text{even}\,,
\end{equation}
where
\begin{equation}\label{3.8}
\left(\overline{
  M_{12}^K}\right)_\text{even}=\lambda_c^{*2}\eta_1S_c+\lambda_t^{*2}\eta_2S_t+2\lambda_c^*\lambda_t^*\eta_3S_{ct}\,,
\end{equation}
and $\hat B_K$ is the well-known non-perturbative factor. Similarly
for $B_d^0-\bar B_d^0$ mixing one has
\begin{equation}\label{3.9}
\left(M_{12}^d\right)_\text{even}=\frac{G_F^2}{12\pi^2}F_{B_d}^2\hat
B_{B_d}m_{B_d}M_{W_L}^2\left(\overline{M_{12}^d}\right)_\text{even}\,,
\end{equation}
where
\begin{equation}\label{3.10}
\left(\overline{
  M_{12}^d}\right)_\text{even}=\left(\lambda_t^{(d)*}\right)^2\eta_BS_t\,.
\end{equation}

In the case of  $B_s^0-\bar B_s^0$ mixing the amplitude $\left(M_{12}^s\right)_\text{even}$ can be obtained
from \eqref{3.9} and \eqref{3.10} by simply replacing $d$ by $s$. It should be
emphasized that $\lambda_i^*$ and not $\lambda_i$ enter these expressions.
Replacing $\lambda_i^*$ erroneously by $\lambda_i$ in \eqref{3.8} would result for
instance in the opposite sign in the CP-violating parameter
$\varepsilon_K$.

\subsection{\boldmath T-odd Sector ($\Delta S=2$)\unboldmath}
\label{subsec:3.2}

The effective Hamiltonians summarizing the contributions of the
mirror fermions and heavy gauge bosons to $\Delta F=2$ transitions
have for the first time been presented in~\cite{Hubisz}. We
confirm the expressions for these Hamiltonians given
in~\cite{Hubisz} but our phenomenological analysis of the
particle-antiparticle mixing presented in Sections~\ref{sec:MFV}
and~\ref{sec:beyond} goes far beyond  the one of these
authors.

\begin{figure}
\center{\epsfig{file=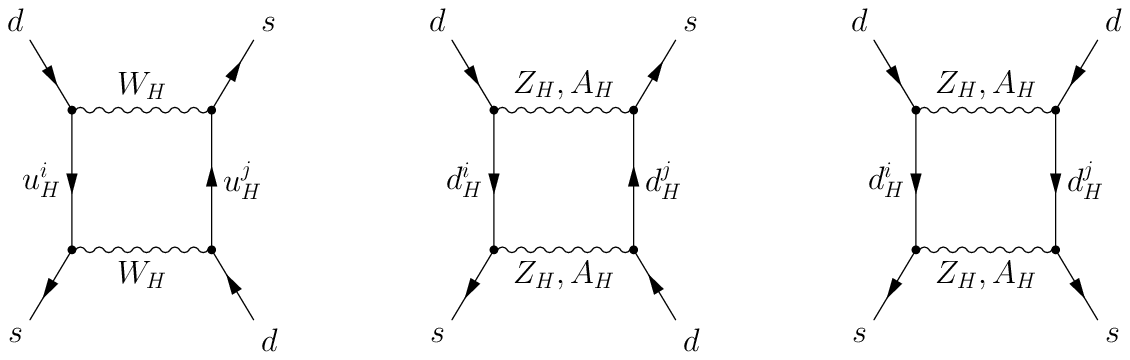,scale=1.1}}
 \caption{\textit{Diagrams contributing to $\Delta F=2$
processes in the T-odd sector.}}\label{fig:mixodd}
\end{figure}

Beginning with $\Delta S=2$ transitions, the contributing diagrams
are shown in Fig.~\ref{fig:mixodd}. The diagrams in which the
gauge bosons run vertically give the same result and bring in a
factor of two. Including the combinatorial factor $1/4$ we find
(the QCD factor $\eta_2$ will be explained in
Section~\ref{subsec:3.5})
\begin{equation}\label{3.11}
\left[\Heff{(\Delta S=2)}\right]_\text{odd}=
\frac{G_F^2}{64\pi^2}M_{W_L}^2\frac{v^2}{f^2}\eta_2\sum_{i,j}\xi_i\xi_j
F_H(z_i,z_j)(\bar sd)_{V-A}(\bar sd)_{V-A}\,,
\end{equation}
where $\xi_i$ have been defined in \eqref{2.12} and $F_H(z_i,z_j)$ with
\begin{equation}\label{3.12}
z_i=\frac{m_{Hi}^2}{M_{W_H}^2}\,,\quad
z_i'=\frac{m_{Hi}^2}{M_{A_H}^2}=z_ia \;\text{ with }
a=\frac{5}{\tan^2{\theta_W}}\qquad (i=1,2,3)\,,
\end{equation}
are given as follows~\cite{Hubisz}
\begin{equation}\label{3.13}
F_H(z_i,z_j)=F(z_i,z_j;W_H)+G(z_i,z_j;Z_H)+A_1(z_i,z_j;Z_H)+A_2(z_i,z_j;Z_H)\,.
\end{equation}

The different contributions correspond to ``$WW$'', ``$ZZ$'', ``$AA$''
and ``$ZA$'' diagrams, respectively. Explicit expressions for the functions $F$, $G$, $A_1$ and $A_2$
are given in Appendix \ref{sec:appB}.

Using the unitarity relation \eqref{2.13} we find then
\begin{eqnarray}\nonumber
\left[\Heff(\Delta S=2)\right]_\text{odd} =
\frac{G_F^2}{64\pi^2}M_{W_L}^2\frac{v^2}{f^2}\eta_2
\left[\xi_2^2R_2(z_1,z_2)+\xi_3^2R_2(z_1,z_3)+2\xi_2\xi_3
  R_3(z_1,z_2,z_3)\right]\\
\cdot(\bar sd)_{V-A}(\bar sd)_{V-A}\,,\qquad\qquad\qquad&&\label{3.14}
\end{eqnarray}
where
\begin{eqnarray}
R_2(z_i,z_j)&=&F_H(z_i,z_i)+F_H(z_j,z_j)-2F_H(z_i,z_j)\label{3.15}\,,\\
R_3(z_1,z_2,z_3)&=&F_H(z_2,z_3)+F_H(z_1,z_1)-F_H(z_1,z_2)-F_H(z_1,z_3)\,.\label{3.16}
\end{eqnarray}

The contribution of the T-odd sector to the off-diagonal
element $M^K_{12}$ in the neutral $K$-meson mass matrix can then be
written similarly to \eqref{3.7} as follows:
\begin{equation}\label{3.17}
\left(M_{12}^K\right)_\text{odd}=\frac{G_F^2}{48\pi^2}F_K^2\hat
B_Km_KM_{W_L}^2\frac{v^2}{f^2}\eta_2\left(\overline{M_{12}^K}\right)_\text{odd}\,,
\end{equation}
where
\begin{equation}\label{3.18}
\left(\overline{M_{12}^K}\right)_\text{odd} =
\xi_2^{*2}R_2(z_1,z_2)+\xi_3^{*2}R_2(z_1,z_3)+2\xi_2^*\xi_3^*R_3(z_1,z_2,z_3)\,.
\end{equation}

\subsection{\boldmath T-odd Sector ($\Delta B=2$\unboldmath)}
\label{subsec:3.3}
It is straightforward to generalize \eqref{3.17} and \eqref{3.18} to
$B_d^0-\bar B_d^0$
and $B_s^0-\bar B_s^0$ mixing. We find for $B_d^0-\bar B_d^0$ mixing
\begin{equation}\label{3.19}
\left(M_{12}^d\right)_\text{odd}=\frac{G_F^2}{48\pi^2}F_{B_d}^2\hat
B_{B_d}m_{B_d}M_{W_L}^2\frac{v^2}{f^2}\eta_{B}\left(\overline{M_{12}^d}\right)_\text{odd}\,,
\end{equation}
where
\begin{equation}\label{3.20}
\left(\overline{M_{12}^d}\right)_\text{odd} =
\left(\xi_2^{(d)*}\right)^2R_2(z_1,z_2)+ \left(\xi_3^{(d)*}\right)^2R_2(z_1,z_3)+2\xi_2^{(d)*}\xi_3^{(d)*}R_3(z_1,z_2,z_3)\,
\end{equation}
with $\xi_i^{(d)}$ defined in \eqref{2.12}.

Finally in the case of $B_s^0-\bar B_s^0$ mixing
\begin{equation}\label{3.21}
\left(M_{12}^s\right)_\text{odd}=\frac{G_F^2}{48\pi^2}F_{B_s}^2\hat
B_{B_s}m_{B_s}M_{W_L}^2\frac{v^2}{f^2}\eta_{B}\left(\overline{M_{12}^s}\right)_\text{odd}\,,
\end{equation}
with $\left(\overline{M_{12}^s}\right)_\text{odd}$ obtained from
$\big(\overline{M_{12}^d}\big)_\text{odd}$ by replacing
$\xi_i^{(d)}$ by $\xi_i^{(s)}$. The QCD factor $\eta_B$ will be
discussed in Section~\ref{subsec:3.5}.

\subsection{Combining T-odd and T-even Sectors}
\label{subsec:3.4}
The final results for $M_{12}^K$, $M_{12}^d$ and $M_{12}^s$ in the LHT model that
govern the analysis of $K^0-\bar K^0$, $B_d^0-\bar B_d^0$ and $B_s^0-\bar B_s^0$ mixing are then
\begin{equation}\label{3.22}
M_{12}^i=\left(M_{12}^i\right)_\text{even}
+\left(M_{12}^i\right)_\text{odd}\qquad (i=K,d,s)\,.
\end{equation}

Let us make a few comments:
\begin{itemize}
\item The new contributions enter both the even and odd
       terms. While the even contributions are dominated by the SM
       part, we will demonstrate in Section~\ref{sec:beyond} that the
       contributions of the even sector to $M_{12}^i$ cannot be generally neglected depending
        on the value of $x_L$ chosen.  
          Due to electroweak precision constraints only certain combinations of $x_L $ and $f$ are allowed~\cite{mH}.
        Therefore the contribution of two $T_+$, represented by
        $\Delta S_{TT}$ in \eqref{TT}, can usually be neglected.
\item In the limit of exactly degenerate mirror quarks the
odd contributions vanish and the LHT model belongs to the class of
models with MFV in which all flavour violating processes are
governed by the CKM matrix and there are no new operators relative
to those present in the SM.\footnote{Strictly speaking the LHT model in the
  limit of degenerate mirror quarks belongs to a subclass of MFV models, the
  so-called Constrained Minimal Flavour Violation (CMFV). In addition to the
  MFV condition that flavour and CP violation is exclusively governed by the
  CKM matrix, in CMFV the structure of low energy operators is the same as in
  the SM. For a detailed discussion on CMFV and MFV we refer to \cite{BBGT}.} As the functions $P_1$ and $P_2$ in
\eqref{3.6} are strictly negative, the new contributions are
strictly positive implying generally lower values of $|V_{td}|$
and $\gamma$ than coming from the SM fits and an enhanced value of
$\Delta M_s$, as already pointed out in~\cite{BPU}. The recent
measurement of $\Delta M_s$ suggesting $\Delta M_s$ possibly
smaller than $(\Delta M_s)_\text{SM}$ puts therefore an important
constraint on the T-even sector unless a rescue comes from the
mirror fermions. We will quantify this in Section~\ref{sec:beyond}.
\item Once the degeneracy of the mirror fermion masses is
removed, two new features appear. First three new complex phases
$\delta_{ij}^d$, generally different from $\delta_\text{CKM}$,
enters the game, with profound consequences for $\varepsilon_K$,
$\Delta \Gamma_q$, $A_\text{SL}^q$, $A_{\rm CP}(B_d \rightarrow
\psi K_S)$, $A_{\rm CP}(B_s \rightarrow \psi \phi)$, $A_{\rm CP}(B
\rightarrow X_{s,d} \gamma)$ and also for $\Delta M_d$ and $\Delta
M_s$ as we will stress below. Equally important, the presence of
new mixing angles $\theta_{ij}^d$, generally different from
$\theta_{ij}$ in the CKM matrix, introduces new flavour violating
interactions leading to the violation of various relations between
$K$, $B_d$, and $B_s$ systems that are characteristic for models
with MFV \cite{UUT,BBGT,MFVrel,Bergmann}. Precisely the violation
of these relations could signal the presence of mirror fermion
contributions.
\end{itemize}

\subsection{QCD Corrections}
\label{subsec:3.5}
QCD corrections to $\Delta F=2$ transitions in the LH model without T-parity and with T-parity have already been discussed in~\cite{BPU} and~\cite{Hubisz}, respectively. Here we will only
summarize the strategy of both papers, that we will follow
throughout our analysis, pointing out the difference between
the QCD corrections in the T-even and T-odd sectors.
\begin{itemize}
\item  Below the thresholds of heavy particles, the QCD
       corrections are at  leading order identical to the ones
       in the SM up to the value of the high energy scale below
       which only SM particles are present in the effective theory.
       This is simply related to the fact that at LO only the
       anomalous dimensions of the involved operators matter~\cite{phen}. As
       the operators present in the LHT model are the same as in the
       SM, the QCD corrections in this approximation can be directly
       obtained from the SM ones, and the same applies to the non-perturbative
       parameters $\hat B_i$ that in fact are identical to the
       ones present in the SM. At NLO, the $\mathcal{O}(\alpha_s)$ corrections at the
       matching scale between the full theory with all heavy particles
       and the effective theory described by the SM will differ
       from the corresponding matching corrections in the SM. The experience from the calculations of such corrections in
       supersymmetric theories, however, shows that they are small due to the
       smallness of $\alpha_s$ at $\mu> M_W$ and that they dominantly serve to
       remove unphysical renormalization scale dependences present
       at LO. Such a calculation is clearly premature at present and
       would only be justified after the discovery of mirror
       fermions, heavy gauge bosons and of $T_\pm$ heavy quarks.
\item  It should also be remarked that a proper calculation of
QCD corrections would require first the knowledge of the full
spectrum of heavy particles involved. In the case of significant
differences between their masses, it could turn out that integrating out all heavy new particles at a
single scale, as done in~\cite{BPU,Hubisz} and here, is not a
satisfactory approximation and the construction of a sequence of effective
theories with a series of thresholds, as done in the SM for scales
below $M_W$, would be necessary~\cite{phen}. Clearly such a difficult task is
premature at present. However, we would like to emphasize the
difference between QCD corrections in the T-even and T-odd
sectors.
\item In the T-odd sector all particles in the loops are
heavy and the structure of the calculation of QCD corrections is
similar to the corresponding calculation of the top contributions
in the SM. Thus for all contributions from mirror fermions we can
use the SM values~\cite{eta2B} as seen in \eqref{3.17},
\eqref{3.19} and \eqref{3.21}
\be
\eta_2=0.57 \pm 0.01\,,\qquad
\eta_B=0.55 \pm 0.01\,.
\ee
\item In the T-even sector also light quarks appear in the
       loops and the calculations of QCD corrections below the
       scale $\mathcal{O}(M_W)$ for charm contributions differ from the one for
       top contributions. In the spirit of the comments made above it is
       a reasonable approximation to use then in this sector~\cite{eta2B}-\cite{eta3}
       \be
\eta_1=1.32\pm 0.32\,,\quad \eta_2=0.57 \pm 0.01\,,\quad \eta_3=0.47\pm 0.05\,,\quad
\eta_B=0.55 \pm 0.01\,.\ee
     The contributions of $T_+$ are absent in the term involving $\eta_1$
       and turn out to be small in the term involving $\eta_3$.
       Consequently $\eta_1$ and  $\eta_3$ are only relevant for the SM
       contributions.
\end{itemize}

\subsection{\boldmath Basic Formulae for $\bm{\varepsilon_K}$ and $\bm{\Delta
    M_i}$ \unboldmath}
\label{subsec:3.6}
In order to study the departures from the SM let us first cast \eqref{3.22}
into
\begin{equation}\label{3.23}
M_{12}^i=\left(M_{12}^i\right)_\text{SM}+\left(M_{12}^i\right)_\text{new}\,,
\end{equation}
with
\begin{eqnarray}
\left(M_{12}^i\right)_\text{new}&=&
\left(M_{12}^i\right)^\text{new}_\text{even}+
\left(M_{12}^i\right)_\text{odd}\label{3.24}\,,\\
 \left(M_{12}^i\right)^\text{new}_\text{even}&=&
 \left(M_{12}^i\right)_\text{even}-
 \left(M_{12}^i\right)_\text{SM}\label{3.25}\,.
\end{eqnarray}
Then the $K_L-K_S$ mass difference is given by
\begin{eqnarray}\nonumber
\Delta M_K&=&2
\left[\text{Re}\left(M_{12}^K\right)_\text{SM}+\text{Re}\left(M_{12}^K\right)_\text{new}\right]\\
&=& \left(\Delta M_K\right)_\text{SM}+\left(\Delta
  M_K\right)_\text{new}\,,\label{3.26}
\end{eqnarray}
and $\varepsilon_K$, neglecting a small contribution involving
$\text{Re}\left(M_{12}^K\right)$, as follows
\begin{eqnarray}\nonumber
\varepsilon_K&=& \frac{e^{i\pi/4}}{\sqrt{2}\left(\Delta
    M_K\right)_\text{exp}}
\left[\text{Im}\left(M_{12}^K\right)_\text{SM}+\text{Im}\left(M_{12}^K\right)_\text{new}\right]\\
&=&
\left(\varepsilon_K\right)_\text{SM}+\left(\varepsilon_K\right)_\text{new}.\label{3.27}
\end{eqnarray}

It should be emphasized that there is no interference
between the SM and new contributions here. They are simply
additive.

We would like to emphasize that this additivity of SM and
new contributions is broken in the case of $\Delta M_d$ and $\Delta M_s$ if the
weak phases of the SM and new contributions differ from each
other. Indeed
\begin{equation}\label{3.28}
\Delta M_q=2\left| \left(M_{12}^q\right)_\text{SM}+
  \left(M_{12}^q\right)_\text{new}\right|\,\qquad (q=d,s)
\end{equation}
and the interference between these two contributions can be
non-zero and not necessarily  constructive. These interferences 
were not discussed in~\cite{Hubisz}, while they will play an important role 
in our numerical analysis.

Let us then write
\begin{eqnarray}
\label{3.29} \left(M_{12}^d\right)_\text{SM}&\equiv&
\left|\left(M_{12}^d\right)_\text{SM}\right|e^{2i\varphi^d_\text{SM}},\qquad
\varphi^d_\text{SM}=\beta\,,\\
\label{3.30} \left(M_{12}^d\right)_\text{new}&\equiv&
\left|\left(M_{12}^d\right)_\text{new}\right|e^{2i\varphi^d_\text{new}}\,,
\end{eqnarray}
and similarly for $M_{12}^s$ with $\varphi^s_\text{SM}=\beta_s - \pi$. Here the phases $\beta$ and $\beta_s$ are
defined through
\begin{equation}\label{3.31}
V_{td}=\left|V_{td}\right|e^{-i\beta}\qquad\text{and}\qquad
V_{ts}=- \left|V_{ts}\right|e^{-i\beta_s}\,,
\end{equation}
with $\beta\simeq 22^\circ$ obtained from UT fits~\cite{UTFIT,CKMFIT} and $\beta_s\simeq -1^\circ$ from the unitarity of
the CKM matrix, its hierarchical structure and its phase
conventions. Consequently we can write
\begin{eqnarray}
\label{3.32}\Delta M_d &=& (\Delta
M_d)_\text{SM}\left|1+h_de^{2i\sigma_d}\right|\equiv (\Delta
M_d)_\text{SM}C_{B_d}\,, \\
\label{3.33}\Delta M_s &=& (\Delta
M_s)_\text{SM}\left|1+h_se^{2i\sigma_s}\right|\equiv (\Delta
M_s)_\text{SM}C_{B_s}\,,
\end{eqnarray}
where we have used the model independent notation of~\cite{NMFV}
and~\cite{UTFIT,Nir}, respectively. Here
\begin{equation}\label{3.34}
h_i=\left|\frac{\left(M_{12}^i\right)_\text{new}}{\left(M_{12}^i\right)_\text{SM}}\right|\,,\qquad
\sigma_i=\varphi^i_\text{new}-\varphi^i_\text{SM}\,.
\end{equation}
We have then
\begin{equation}\label{3.35}
\frac{\Delta M_d}{\Delta M_s}=\frac{m_{B_d}}{m_{B_s}}\frac{\hat
  B_{B_d}}{\hat
  B_{B_s}}\frac{F_{B_d}^2}{F_{B_s}^2}\left|\frac{V_{td}}{V_{ts}}\right|^2\frac{C_{B_d}}{C_{B_s}}\,,
\end{equation}
and the MFV relation between ${\Delta M_d}/{\Delta M_s}$ and
$\left|{V_{td}}/{V_{ts}}\right|^2$ is violated if $C_{B_d}\ne
C_{B_s}$. We will investigate this violation in
Section~\ref{sec:beyond}.

\subsection{$\bm{A_{\rm CP}^\text{mix}(B_d \rightarrow \psi K_S)}$
  and $\bm{A_{\rm CP}^\text{mix}(B_s \rightarrow \psi\phi)}$}
\label{subsec:3.7}

The next to be considered on our list are the mixing induced
CP asymmetries in $B_d^0 \rightarrow \psi K_S$ and $B_s^0 \rightarrow \psi\phi$ decays, that within the SM and
MFV models provide the measurements of the phases $\beta$ and $\beta_s$,
respectively, without essentially any theoretical
uncertainty. This clean character remains true within the
LHT model since
\begin{itemize}
\item there are no new tree level contributions to the decay
       amplitudes for $B_d^0 \rightarrow \psi K_S$ and $B_s^0 \rightarrow \psi\phi$ as they are forbidden by T-parity.
\item  there are no new operators beyond the ones with the $(V-
A)\otimes(V-A)$ structure present in the SM, implying that all non-perturbative uncertainties present in $M^i_{12}$ will cancel in
evaluating the CP asymmetries.
\end{itemize}
Denoting then as in~\cite{NMFV} and~\cite{UTFIT,Nir}
\begin{equation}\label{3.36}
1+h_ie^{2i\sigma_i}=\left|1+h_ie^{2i\sigma_i}\right|e^{2i\varphi_{B_i}}\equiv
C_{B_i}e^{2i\varphi_{B_i}} \,,
\end{equation}
one finds the formulae for the coefficients $S_{\psi K_S}$ and
$S_{\psi\phi}$ of $\sin{(\Delta M_dt)}$ and $\sin{(\Delta M_st)}$,
respectively, in the time dependent asymmetries in question
\begin{eqnarray}
\label{3.37}
S_{\psi K_S}&=&-\eta_{\psi K_S}\sin{(2\beta+2\varphi_{B_d})}=\sin{(2\beta+2\varphi_{B_d})}\,,\\
\label{3.38}
S_{\psi\phi}&=&-\eta_{\psi \phi}\sin{(2\beta_s+2\varphi_{B_s})}=\sin{(2|\beta_s|-2\varphi_{B_s})}\,,
\end{eqnarray}
where $\eta_{\psi K_S}$ and $\eta_{\psi\phi}$ are the CP parities of
the final states. We set $\eta_{\psi\phi}=+1$. Thus in the
presence of new contributions with $\sigma_i\ne 0,\pi/2$, or
equivalently $\varphi^i_\text{new}\ne\varphi^i_\text{SM}$,
$S_{\psi K_S}$ and $S_{\psi\phi}$ will not measure the phases
$\beta$ and $\beta_s$ but $(\beta+\varphi_{B_d})$ and
$(|\beta_s|-\varphi_{B_s})$, respectively. We will return to
investigate this effect numerically in Section~\ref{sec:beyond}.
Note that it is $-\varphi_{B_s}$ and not $+\varphi_{B_s}$ that
enters~(\ref{3.38})~\cite{BBGT}.

\subsection{\boldmath${\Delta \Gamma_q}$ and ${A_\text{SL}^q}$\unboldmath}
\label{subsec:3.8}
The last quantities we will consider in this section are the width difference
$\Delta \Gamma_q$ and the semileptonic CP asymmetry $A_\text{SL}^q$, defined respectively as
\begin{gather}
\Delta \Gamma_q = \Gamma_L^q - \Gamma_H^q\,,
\label{3.39a}\\
A_\text{SL}^q=\frac{\Gamma(\bar
  B_q^0\rightarrow\ell^+X)-\Gamma(B_q^0\rightarrow\ell^-X)}{\Gamma(\bar
  B_q^0\rightarrow\ell^+X)+\Gamma(B_q^0\rightarrow\ell^-X)}\,,
\label{3.39b}
\end{gather}
with $q=d,\,s$ and the light and heavy mass eigenstates given by
\be
|B_q^{L,H} \rangle = \frac{1}{\sqrt{1+|q/p|_q^2}}\,\left( |B^0_q \rangle \pm
  \left(\frac{q}{p}\right)_q |\bar B^0_q \rangle \right )\,.
\label{3.40}
\ee

Width difference and semileptonic CP asymmetry are obtained by diagonalizing
the $2 \times 2$ Hamiltonian which describes the $B^0_q - \bar B^0_q$ systems.
Neglecting terms of $\mathcal{O}(m_b^4/m_t^4)$, they can simply be written as
\begin{gather}
\Delta \Gamma_q = -\Delta M_q\,\text{Re}\left
  (\frac{\Gamma_{12}^q}{M_{12}^q} \right)\,,\label{3.41a}\\
A_\text{SL}^q = -2\left(\left| \frac{q}{p}\right|_q-1\right) = \text{Im}\left
  (\frac{\Gamma_{12}^q}{M_{12}^q} \right)\,,\label{3.41b}
\end{gather}
where $\Gamma_{12}^q$ is the absorptive part of the $B^0_q - \bar B^0_q$
amplitude.
Theoretical predictions of both $\Delta \Gamma_q$ and $A_\text{SL}^q$,
therefore, require the calculation of the off-diagonal matrix element
$\Gamma_{12}^q$.

Important theoretical improvements have been achieved thanks to advances in
lattice studies of $\Delta B=2$ four-fermion operators~\cite{Becirevic:2001xt}-\cite{Aoki:2003xb} and to the NLO
perturbative calculations of the corresponding Wilson
coefficients~\cite{noi}-\cite{Beneke:2003az}.
From slight updates to the theoretical analysis performed in~\cite{noi} we find
\begin{gather}
\text{Re}\left
  (\frac{\Gamma_{12}^d}{M_{12}^d} \right) = -(3.0 \pm 1.0)\cdot10^{-3}\,,\qquad
\text{Re}\left
  (\frac{\Gamma_{12}^s}{M_{12}^s} \right) = -(2.6 \pm 1.0)\cdot10^{-3} \,,\label{eq:r2}\\
\text{Im}\left
  (\frac{\Gamma_{12}^d}{M_{12}^d} \right) = -(6.4 \pm 1.4)\cdot
10^{-4}\,,\qquad \text{Im}\left
  (\frac{\Gamma_{12}^s}{M_{12}^s} \right) = (2.6 \pm 0.5)\cdot 10^{-5}\,,\label{eq:r1}
\end{gather}
which, combined with the experimental values of lifetimes and mass
differences, yield
\begin{gather}
\frac{\Delta \Gamma_{d}}{\Gamma_d} = (2.3 \pm 0.8)\cdot10^{-3}\,,\qquad
\frac{\Delta \Gamma_{s}}{\Gamma_s} = (6.7 \pm 2.7)\cdot 10^{-2} \,,\label{eq:r4}\\
A_\text{SL}^d = -(6.4 \pm 1.4)\cdot 10^{-4}\,,\qquad
A_\text{SL}^s = (2.6 \pm 0.5) \cdot 10^{-5}\,.\label{eq:r3}
\end{gather}
We note that the theoretical prediction for
$\text{Re}(\Gamma_{12}^s/M_{12}^s)$ 
obtained in~\cite{noi} and updated in~(\ref{eq:r2}) is smaller than the value found in \cite{Beneke:1998sy}. 
This difference is mainly due to the contribution of
$\mathcal{O}(1/m_b^4)$ in the Heavy Quark Expansion (HQE), which in \cite{Beneke:1998sy} is
wholly estimated in the vacuum saturation approximation (VSA), while in
\cite{noi} the matrix elements of two dimension-seven operators are expressed
in terms of those calculated on the lattice. Being the $\ord(1/m_b^4)$
contribution important, it is interesting to estimate the size of the
$\ord(1/m_b^5)$ terms. A perturbative calculation of the corresponding
Wilson coefficients is now in progress \cite{Lenz}.

On the experimental side new relevant measurements exist. The
averaged experimental results and limits read~\cite{BBpage}
\begin{gather}
\frac{\Delta \Gamma_{d}}{\Gamma_d} = 0.009 \pm 0.037 \,,\qquad
\frac{\Delta \Gamma_{s}}{\Gamma_s} = 0.31^{+0.10}_{-0.11}\,,
\label{3.43b}\\
A_\text{SL}^d=-(0.0030\pm 0.0078)\,.\label{3.43a}
\end{gather}
The comparisons are not yet conclusive, due to still large
experimental uncertainties, whose reduction is certainly being looked
forward to.

The great interest in confirming or not the SM predictions comes from the
sensitivity of these observables to new physics.
In the presence of new phases beyond the CKM one, whose effect on  $M_{12}^q$
follows directly from~(\ref{3.36}), one finds   
\begin{gather}
\frac{\Delta \Gamma_q}{\Gamma_q} = -\,\left(\frac{\Delta
    M_q}{\Gamma_q}\right)^\text{exp}\,\left[\text{Re}\left(\frac{\Gamma^q_{12}}{M^q_{12}}\right)^\text{SM}\frac{\cos{2\varphi_{B_q}}}{C_{B_q}}-
\text{Im}\left(\frac{\Gamma^q_{12}}{M^q_{12}}\right)^\text{SM}\frac{\sin{2\varphi_{B_q}}}{C_{B_q}}\right]\,,\label{3.44a}\\
A_\text{SL}^q=\text{Im}\left(\frac{\Gamma^q_{12}}{M^q_{12}}\right)^\text{SM}\frac{\cos{2\varphi_{B_q}}}{C_{B_q}}-
\text{Re}\left(\frac{\Gamma^q_{12}}{M^q_{12}}\right)^\text{SM}\frac{\sin{2\varphi_{B_q}}}{C_{B_q}}\,.\label{3.44b}
\end{gather}

It is important to note that with $\text{Re}(\Gamma_{12}^s/M_{12}^s) \gg
\text{Im}(\Gamma_{12}^s/M_{12}^s)$, even a
small $\varphi_{B_s}$ can induce an order of magnitude enhancement of
$A_\text{SL}^s$ relative to the SM.
On the other hand, a non-vanishing $\varphi_{B_q}$ would result in a
suppression of $\Delta \Gamma_q/\Gamma_q$, thus increasing the discrepancy
with the experimental average in the $q=s$ case. We note, however, that the new preliminary experimental average $\Delta
\Gamma_s/\Gamma_s= 0.14 \pm 0.06$ \cite{FPCP} is lower than the previous
one, thus reducing significantly the discrepancy with the SM theoretical
prediction in (\ref{eq:r4}).
These topics have been extensively discussed in the recent
literature~\cite{BBGT,UTFIT,Ligeti,Grossman:2006ce}. 
In \cite{Ligeti} a correlation between $A_{\rm SL}^s$ and $S_{\psi\phi}$ 
has been pointed out and 
in~\cite{BBGT} some correlations have been derived in order to determine 
the ratio $\Delta M_q/(\Delta M_q)_\text{SM}$ in a
theoretically clean way.
They read 
\begin{gather}
\frac{\Delta M_q}{(\Delta M_q)_\text{SM}}= \left |
 \text{Re}\left(\frac{\Gamma^q_{12}}{M^q_{12}}\right)^\text{SM}\right|
\frac{\sin 2
  \varphi_{B_q}}{A^q_\text{SL}}+\text{Im}\left(\frac{\Gamma^q_{12}}{M^q_{12}}\right)^\text{SM}\frac{\cos{2\varphi_{B_q}}}{A^q_\text{SL} }\,,\label{eq:DMratio}\\
\frac{\Delta M_q}{(\Delta M_q)_\text{SM}} = 
-\left(\frac{\Delta M_q}{\Delta \Gamma_q}\right) \text{Re}\left(\frac{\Gamma^q_{12}}{M^q_{12}}\right)^\text{SM} \cos 2\varphi_{B_q}\,,
\label{eq:DMqr}
\end{gather}
with $\varphi_{B_q}$ to be extracted from $S_{\psi\phi}$ and $S_{\psi K_S}$ 
for $q=s$ and $q=d$, respectively. In the case of $q=s$, the second term in 
(\ref{eq:DMratio}) can be safely neglected.
It will be interesting to consider these correlations within the LHT model once
the experimental uncertainties have been significantly reduced.

\subsection{Summary}
\label{subsec:3.9}

In this section we have calculated the $\mathcal{O}({v^2/f^2})$
corrections to the amplitudes $M_{12}^K$, $M_{12}^d$ and
$M_{12}^s$ in the LHT model confirming the results of~\cite{Hubisz}.
We have then given formulae for $\Delta M_K$,
$\Delta M_d$, $\Delta M_s$, $\varepsilon_K$, $S_{\psi K_S}$,
$S_{\psi\phi}$, $\Delta \Gamma_q$ and $A^q_\text{SL}$ in a form
suitable for the study of the size of the new LHT contribution.
The numerical analysis of these observables will be presented in
Section~\ref{sec:beyond}.

\newsection{\boldmath $B \rightarrow X_s \gamma$ in the LHT Model \unboldmath}
\label{sec:bsg}
\subsection{Preliminaries}
\label{subsec:4.1}

One of the most popular decays used to constrain new physics
contributions is the $B \rightarrow X_s \gamma$ decay for which
the measured branching ratio~\cite{BBpage} 
\be Br(B \rightarrow
X_s \gamma)_\text{exp} = (3.52 \pm 0.30) \cdot 10^{-4}
\label{eq:bsgexp} \ee 
agrees well with the SM NLO
prediction~\cite{bsgSM1,bsgSM2}  
\be Br(B \rightarrow X_s
\gamma)_\text{SM} = (3.33 \pm 0.29) \cdot 10^{-4}\,,
\label{eq:bsgSM} \ee
both given for $E_\gamma>1.6\gev$ and the SM prediction for
$m_c(m_c)/m_b^{1S} = 0.26$. 
For $Br(B\to X_d\gamma)$, instead, the SM prediction is in the
ballpark of $1.5\cdot 10^{-5}$.

One should emphasize that within the SM this decay is governed by
the already well determined CKM element $|V_{ts}|$  so that
dominant uncertainties in~(\ref{eq:bsgSM}) result from the
truncation of the QCD perturbative series and the value of
$m_c(\mu)$ that enters the branching ratio first at the NLO level.
A very difficult NNLO calculation, very recently completed~\cite{bsgSM2}, 
reduced the error in~(\ref{eq:bsgSM}) significantly below $10$\%:
$(3.15\pm0.23)\cdot 10^{-4}$.

The effective Hamiltonian relevant for this decay within the SM is
given as follows
\be
\Heff^\text{SM}(\bar b \rightarrow \bar s
\gamma) = - \GF V_{ts} V_{tb}^* \left [ \sum_{i=1}^6 C_i(\mu_b)Q_i
+ C_{7\gamma}(\mu_b)Q_{7\gamma} + C_{8G}(\mu_b)Q_{8G}  \right ],
\label{eq:Heffbsg}
\ee
where $Q_i$ are four-quark operators,
$Q_{7\gamma}$ is the magnetic photon penguin operator and $Q_{8G}$
the magnetic gluon penguin operator. The explicit expression for
the branching ratio $Br(B \rightarrow X_s \gamma)$ resulting
from~(\ref{eq:Heffbsg}) is very complicated and we will not present it
here. It can be found for instance in~\cite{bsgSM1}.

For our purposes it is sufficient to know that in the LO
approximation the Wilson coefficients $C_{7 \gamma}$ and $C_{8G}$ are given
at the renormalization scale $\mu_W=\mathcal{O}(M_W)$ as follows
\be
C_{7\gamma}^0(\mu_W) = - \dfrac{1}{2} D_0'(x_t)\,, \qquad C_{8G}^0(\mu_W) = -
\dfrac{1}{2} E_0'(x_t)\,,
\label{eq:C7g0C8G0}
\ee
with the explicit expressions for $D_0'(x_t)$ and $E_0'(x_t)$ given in Appendix
\ref{sec:appB}.

In view of the importance of QCD corrections in this decay
we will include these corrections at NLO in the SM
part, but only at LO in the new contributions. This
amounts to including only corrections to the renormalization
of the operators in the LHT part and eventually to increase
the scale $\mu_W$ to $\mu \approx 500 \gev$ at which the new particles are
integrated out. As the dominant QCD corrections to $Br(B \rightarrow X_s
\gamma)$ come anyway from the renormalization group evolution from $\mu_W$ down to $\mu_b=\mathcal{O}(m_b)$ and
the matrix elements of the operators $Q_2$ and $Q_{7\gamma}$ at $\mu_b$, these
dominant corrections are common to the SM and LHT parts. 

Within the LO approximation the new physics contributions to
$B \rightarrow X_s \gamma$ enter only through the modifications of the
following two combinations
\be
T_{D'}^\text{SM} \equiv \lambda_t^{(s)} D_0'(x_t)\,, \qquad
T_{E'}^\text{SM} \equiv \lambda_t^{(s)} E_0'(x_t)\,,
\label{eq:TDTE}
\ee
with the CKM factor $\lambda_t^{(s)}=V_{ts}V_{tb}^*$.

\subsection{T-even Sector}
\label{subsec:4.2}

\begin{figure}
\center{\epsfig{file=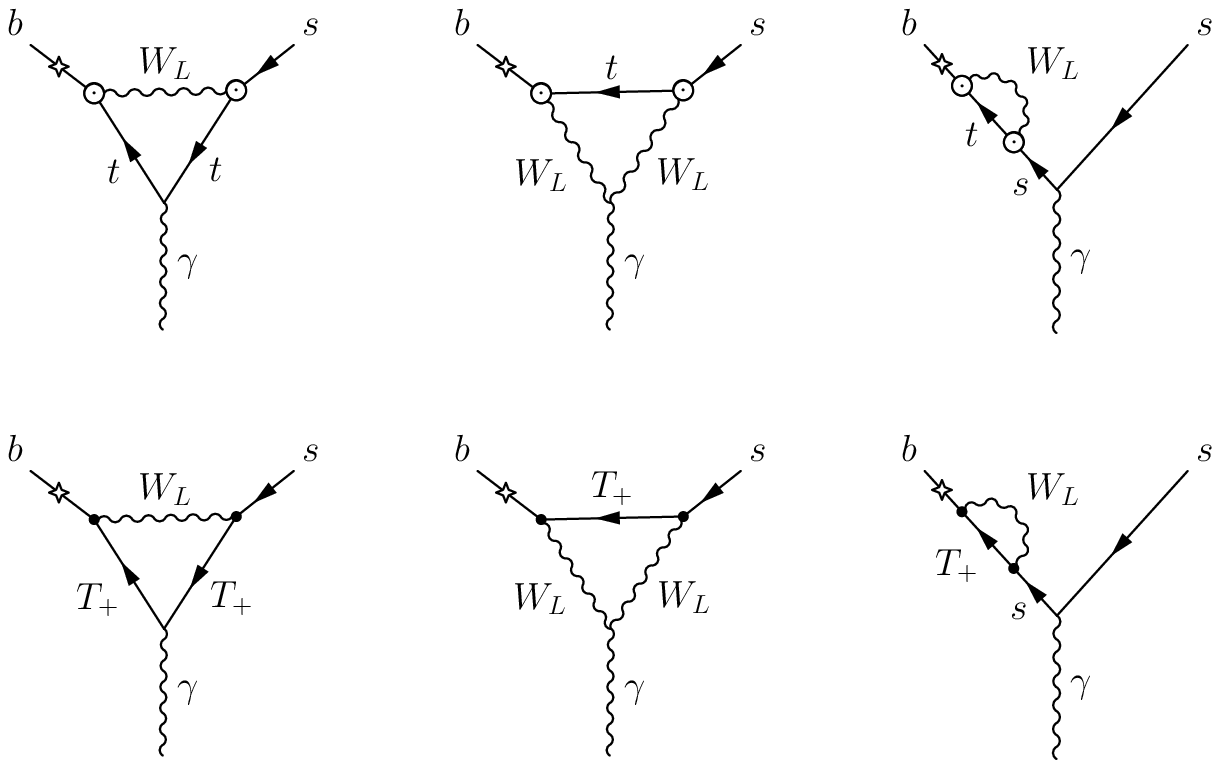}}
 \caption{\textit{New diagrams contributing to
$B\to X_s\gamma$ in the T-even sector.}}\label{fig:bsgeven}
\end{figure}

The first calculation of the $B \rightarrow X_s \gamma$ decay
within the LH model has been done within the Littlest Higgs model
without T-parity in~\cite{bsgl2h}. We have confirmed this result.
Specializing it to the LHT model leaves
 at $\mathcal{O}(v^2/f^2)$ only the contributions shown in
Fig.~\ref{fig:bsgeven}. The result can be directly obtained by
changing the arguments in $D_0'(x_t)$ and $E_0'(x_t)$. We find
then \bea
T_{D'}^\text{even}&=&\lambda_t^{(s)}\left[D_0'(x_t)+\dfrac{v^2}{f^2}x_L^2
\big(D_0'(x_T)-D_0'(x_t)\big)\right]\,,\label{eq:TDeven}\\
T_{E'}^\text{even}&=&\lambda_t^{(s)}\left[E_0'(x_t)+\dfrac{v^2}{f^2}x_L^2
\big(E_0'(x_T)-E_0'(x_t)\big)\right]\,,\label{eq:TEeven}
\eea
with
$x_t$ and $x_T$ defined in~(\ref{eq:xtxT}).

The calculation for the $B \rightarrow X_d \gamma$ decay is completely
analogous and the corresponding T-even contributions can be obtained from
(\ref{eq:TDeven}) and (\ref{eq:TEeven}) with the replacement $s \rightarrow d$. 

\subsection{T-odd Sector}
\label{subsec:4.3}

\begin{figure}
\center{\epsfig{file=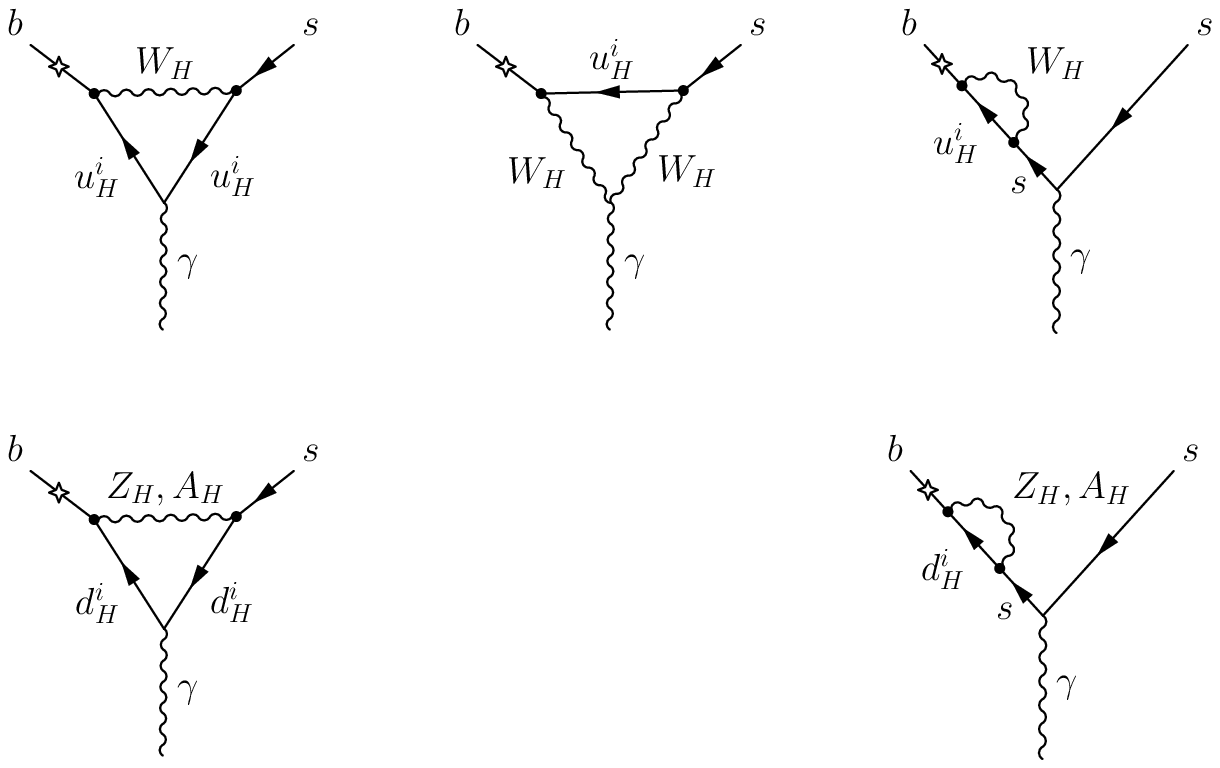}}
 \caption{\textit{Diagrams contributing
to $B\to X_s\gamma$ in the T-odd sector.}}\label{fig:bsgodd}
\end{figure}

The diagrams contributing at $\mathcal{O}(v^2/f^2)$ in this sector
are shown in Fig.~\ref{fig:bsgodd}. The results for these diagrams
can be easily obtained from $D_0'(x_t)$ and $E_0'(x_t)$ as
follows. The contributions from $W_H^{\pm}$ can be found directly
as in the even sector. The contributions of $A_H$ and $Z_H$ can be
on the other hand obtained from $E_0'(x_t)$ as, similarly to the
gluon penguin, they do not contain triple weak gauge boson
vertices.

We find first using the unitarity of the $V_{H_d}$ matrix \bea
T_{D'}^\text{odd}&=&\dfrac{1}{4}\dfrac{v^2}{f^2}\left[\xi_2^{(s)}
\big(D'_\text{odd}(z_2)-D'_\text{odd}(z_1)\big)+\xi_3^{(s)}
\big(D'_\text{odd}(z_3)-D'_\text{odd}(z_1)\big)\right]\,,\label{eq:TDodd}\\
T_{E'}^\text{odd}&=&\dfrac{1}{4}\dfrac{v^2}{f^2}\left
[\xi_2^{(s)}\big(E'_\text{odd}(z_2)-E'_\text{odd}(z_1)\big)+\xi_3^{(s)}
\big(E'_\text{odd}(z_3)-E'_\text{odd}(z_1)\big)\right]\,.\label{eq:TEodd}
\eea
A straightforward calculation gives then
\bea
D'_\text{odd}(z_i)&=&D_0'(z_i)-\dfrac{1}{6}E_0'(z_i)-\dfrac{1}{30}E_0'(z_i')\,,\label{eq:Dodd}\\
E'_\text{odd}(z_i)&=&E_0'(z_i)+\dfrac{1}{2}E_0'(z_i)+\dfrac{1}{10}E_0'(z_i')\,,\label{eq:Eodd}
\eea
where the three contributions correspond to $W_H$, $Z_H$ and $A_H$ exchanges, respectively.
The variables $z_i$ and $z_i'$ are defined in~(\ref{3.12}).

Similarly to the T-even sector, the T-odd contributions to the $B \rightarrow X_d \gamma$
decay can be obtained in a trivial way, with the replacement $s \rightarrow d$
in (\ref{eq:TDodd}) and (\ref{eq:TEodd}) . 

\subsection{\boldmath CP Asymmetry in $B\to X_{s,d}\gamma$ \unboldmath}
\label{subsec:4.4}

In view of the new weak phases present in the LHT model, of particular
interest is the direct CP asymmetry in $B\to X_s\gamma$ that due to a very
small phase of $V_{ts}$ is about $0.5\%$ in the SM. In the case of $B\to X_d\gamma$, the corresponding asymmetry is
governed in the SM by the phase $\gamma+\beta$ and is about
$-10\%$. Consequently, it will be harder to see new physics in this
case unless the experiment shows an opposite sign. Defining
\be
C_{7\gamma}(m_b)=-|C_{7\gamma}(m_b)| e^{i\phi_7}\,, \qquad
C_{8G}(m_b)=-|C_{8G}(m_b)| e^{i\phi_8}\,,
\ee
and using the formulae of~\cite{KN} it is straightforward to calculate
the CP asymmetries in question.
We recall that in the SM $\phi_7=\phi_8=0$.

\subsection{Summary}
\label{subsec:4.5}
In this section we have calculated, for the first time, the
$\mathcal{O}(v^2/f^2)$ corrections to the $B
\rightarrow X_s \gamma$
decay in the LHT model. The final results can be summarized by
\be
T_{D'}=T_{D'}^\text{even}+T_{D'}^\text{odd}\,, \qquad
T_{E'}=T_{E'}^\text{even}+T_{E'}^\text{odd}\,,
\ee
with the
various terms given in~(\ref{eq:TDeven})--(\ref{eq:TEodd}). The numerical analysis of
the branching ratios and the CP asymmetries in question 
will be given in Section~\ref{sec:beyond}.

 Our result for $T_{E'}$
can also be used for the $b \rightarrow s \,g$ decay, but in view of very
large theoretical uncertainties in the corresponding
branching ratio we will not consider it here.

\newsection{Strategy and Goals}\label{sec:goals}

\begin{figure}
\center{\epsfig{file=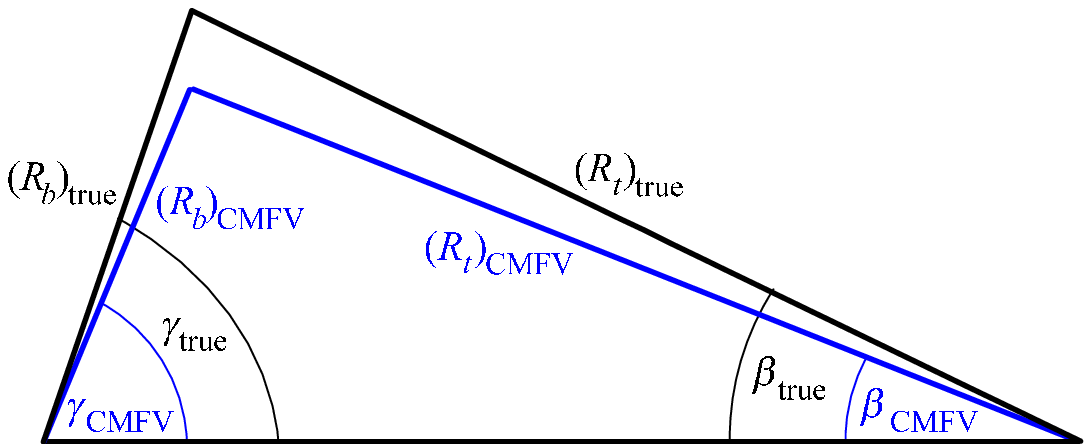}}
\caption{\textit{Reference unitarity triangle and universal unitarity
  triangle \cite{BBGT}.}}
\label{fig:RUT-UUT}
\end{figure}

In what follows it will be useful to recall the unitarity triangle
shown in Fig.~\ref{fig:RUT-UUT}
with $R_b$ and $R_t$ given as follows
\begin{gather}
R_b = \frac{| V_{ud}^{}V^*_{ub}|}{| V_{cd}^{}V^*_{cb}|},
\qquad
\label{2.95}
R_t = \frac{| V_{td}^{}V^*_{tb}|}{| V_{cd}^{}V^*_{cb}|} .
\end{gather}

Using $R_b$ and $\gamma$ determined in tree level decays one can
construct the so-called reference unitarity triangle (RUT)
\cite{refut} with $R_b$ and $\gamma$ independent of new physics
contributions and denoted therefore by $(R_b)_\text{true}$ and $\gamma_\text{true}$. On the other hand using the MFV relations
\cite{BBGT}
\begin{gather}\label{R1}
\sin 2\beta= S_{\psi K_S}\equiv \sin 2\beta_\text{MFV}\,,\\
R_t
\label{RRt} = 0.923~\left[\frac{\xi}{1.23}\right]
\sqrt{\frac{17.4/\text{ps}}{\Delta M_s}} \sqrt{\frac{\Delta
M_d}{0.507/\text{ps}}}\equiv (R_t)_\text{MFV}\,,
\end{gather}
where  \cite{Hashimoto} \be\label{xi} \xi = \frac{\sqrt{\hat
B_{B_s}}F_{B_s} }{ \sqrt{\hat B_{B_d}}F_{B_d}}=1.23\pm 0.06, \ee
allows to construct the UUT \cite{UUT}. 
The two triangles are
related through
\be\label{VUBG}
R_b=\sqrt{1+R_t^2-2
R_t\cos\beta},\qquad
\cot\gamma=\frac{1-R_t\cos\beta}{R_t\sin\beta}\,,
\ee
and the
violation of these relations would in the context of the LHT model
signal the presence of new flavour and CP-violating interactions.
Indeed the low energy operator structure in the LHT model is the
same as in the SM and rescue from new operators cannot be
expected.

A detailed test of the relations in \eqref{VUBG} is presently not
possible in view of sizable theoretical and experimental
uncertainties and the fact that the two triangles do not differ by
much from each other as seen in Fig.~\ref{fig:RUT-UUT}. Yet, if one desperately looks for some differences
between these two triangles one finds that the ``true'' values of
various parameters extracted from the RUT differ from the
corresponding MFV values \cite{BBGT,UTFIT,BFRS05,Branco}
\be\label{P1}
\beta_{\rm true} >
\beta_{\rm MFV}\,, \quad \gamma_{\rm true} > \gamma_{\rm
MFV}\,, \quad (R_t)_{\rm true} > (R_t)_{\rm MFV}\,, \quad (R_b)_{\rm true} > (R_b)_{\rm MFV}\,.
\ee
In particular, there is a tension between
the MFV value of $\sin 2\beta$ and the one indicated by the true value of
$R_b$, as discussed in detail in \cite{BBGT}.

Moreover, the measured value of $\Delta M_s$
\cite{CDFnew}
\be\label{CDF}
\Delta M_{s}=(17.33^{+0.42}_{-0.21}\pm
0.07)/{\rm ps}
\ee
turned out to be surprisingly below the SM
predictions obtained from other constraints \cite{UTFIT,CKMFIT}
\be\label{DMsSM} \left(\Delta
  M_s\right)^\text{SM}_\text{UTfit}=(21.5\pm2.6)/\text{ps},\qquad
\left(\Delta
  M_s\right)^\text{SM}_\text{CKMfitter}=\left(21.7^{+5.9}_{-4.2}\right)/\text{ps}.
\ee
The slight tension between \eqref{CDF} and \eqref{DMsSM} is
not yet significant as the non-perturbative uncertainties are large
but it appears that
\be\label{P2}
\Delta M_s < (\Delta M_s)_{\rm SM}
\ee
could be favoured, and a confident verification of \eqref{P2} would be important.
As recently demonstrated in \cite{BB}, in fact, in 
models with 
constrained MFV, in which the flavour violation is governed entirely by
the SM Yukawa couplings and to a very
good approximation there are no new operators beyond those present in the
SM, $\Delta M_s\ge (\Delta M_s)_{\rm SM}$.

Our three goals for the next two Sections are then as follows:
\begin{enumerate}
\item We will consider the MFV limit of the LHT model in which the
  problems in
(\ref{P1}) and (\ref{P2}) cannot be solved.
\item We will investigate whether the problems listed in (\ref{P1}) and (\ref{P2})
can be solved within the LHT model with the help of new flavour
and CP-violating interactions encoded in the matrix $V_{Hd}$
by choosing a special pattern of the mirror fermion mass spectrum.
\end{enumerate}
Once new CP-violating phases are present, the CP asymmetries
$A_{\rm CP}(B_s\to\psi\phi)$, $A_{\rm SL}^q$ and
$A_{\rm CP}(B\to X_s\gamma)$ can be significantly larger than in
the SM. This brings us to our third goal:
\begin{enumerate}\setcounter{enumi}{2}
\item We will look for interesting benchmark scenarios for the matrix $V_{Hd}$
  and for the mass spectrum of mirror fermions 
in which large enhancements
of $A_{\rm CP}(B_s\to\psi\phi)$, $A_{\rm SL}^q$ and
$A_{\rm CP}(B\to X_s\gamma)$ over the SM values are possible being 
still consistent with all other
constraints. We will also investigate the implications for $\Delta\Gamma_q/\Gamma_q$.
\end{enumerate}

\newsection{Benchmark Scenarios for New Parameters}
\label{sec:MFV}
\subsection{Preliminaries}
\label{subsec:6.1}

In what follows we will consider several scenarios for the
structure of the $V_{Hd}$ matrix and the mass spectrum of mirror
fermions with the hope to gain a global view about the
possible signatures of mirror fermions in the processes
considered and of $T_+$ present in the T-even contributions.
In all these scenarios we will set to zero the phases $\delta_{12}^d$ and
  $\delta_{23}^d$ of $V_{Hd}$,
whose presence was overlooked in \cite{Hubisz} and in the first version of the
present work while has been first pointed out in \cite{SHORT}.
This assumption is quite reasonable, since the impact of the additional
  two phases is numerically small and would not change qualitatively our
  results.

The most interesting scenarios in the model in question will be those
in which the mixing matrix $V_{Hd}$ differs significantly 
from $V_\text{CKM}$ and
has a non-vanishing complex phase $\delta_{13}^d$. As now the number of
parameters increases significantly, it is essential to
determine the CKM parameters from tree level decays.
The left-over room for new physics
contributions will depend on the outcome of this
determination.

Now for given values of $|V_{ub}|$, $|V_{cb}|$ and $|V_{us}|$  and the angle $\gamma$ in the CKM
unitarity triangle, the true phases of the $V_{td}$ and $V_{ts}$
couplings are determined and the asymmetries $S_{\psi K_s}$ and $S_{\psi \phi}$ in~(\ref{3.37}) and~(\ref{3.38}), respectively, can be predicted by setting
first $\varphi_{B_d}$ and $\varphi_{B_s}$ to zero. Similarly, $\eps_K$, $\Delta
M_K$, $\Delta M_d$, $\Delta M_s$ and $Br(B \rightarrow X_s \gamma)$ can be
predicted and compared with the experiments, possibly revealing
the need for new physics contributions, as discussed in previous sections.

In the next section we will be  primarily interested in achieving
the three  goals listed in Section 5. Moreover, it will be
interesting to see how the MFV correlations between $K^0$,
$B^0_d$ and $B_s^0$ systems are modified when new sources of
flavour and CP violation are present. Effectively, such
modifications can be studied by introducing effective one-loop
 functions $(S_i)_{\rm eff}$
\be
(S_i)_{\rm eff}=S_0(x_t)~ C_i~ e^{2i\varphi_i}, \qquad i=K,B_d,B_s
\ee
with $C_i$ and $\varphi_i$ already defined in \eqref{3.36}. In MFV
\be\label{eq:KdsMFV} 
C_K=C_{B_d}=C_{B_s}\,, \qquad
\varphi_K=\varphi_{B_d}=\varphi_{B_s}=0\,,
\ee
but as we will see
below, this is generally not the case in the LHT model. In
particular we will investigate the violation of the MFV
relation between $\Delta M_d/\Delta M_s$ and $|V_{td}/V_{ts}|$
obtained for $C_{B_d}=C_{B_s}$ in \eqref{3.35}.

It is not a purpose of our numerical analysis of the next section to 
consider the full space of parameters but rather to have a closer look 
at a number of scenarios in which some of the problems listed above 
can be simply addressed. Not all the scenarios listed below solve the
problems in question and some of them give results that are very close to 
the SM predictions, 
but we found at least one scenario (S4) in which all our goals have been 
achieved. In this scenario, the $V_{Hd}$ matrix takes a hierarchical structure that
is very different from the structure of the CKM matrix.

\subsection{Different Scenarios}
\label{Scenarios}
Here we just list the scenarios in question:

\vspace{0.2cm}
{\bf{Scenario 1:}}

\noindent
In this scenario the mirror fermions will be degenerate in mass 
\be
m_{H1}=m_{H2}=m_{H3}
\ee
and only the T-even sector will contribute. This is the MFV limit of the LHT
model.

\vspace{0.2cm}
{\bf Scenario 2:} 

\noindent
In this scenario the mirror
fermions are not degenerate in mass and
\be
V_{Hd}=V_\text{CKM}\,.
\label{eq:VHdVCKM}
\ee
In this case there are no contributions of mirror fermions
to $D^0 - \bar D^0$ mixing and flavour violating $D$ meson decays, and
\be
\xi_2^{(q)}=\lambda_c^{(q)}\,, \qquad \xi_3^{(q)}=\lambda_t^{(q)}\,,
\label{eq:xi2xi3}
\ee
with $q=d,s$ and no index $q$ in the $K$ system.

\vspace{0.2cm}
{\bf Scenario 3:}

\noindent
 In this scenario we will choose a linear spectrum for mirror fermions
\be\label{S3}
m_{H1}=400\gev, \qquad  m_{H2}=500\gev, \qquad m_{H3}=600\gev
\ee
but an otherwise arbitrary matrix $V_{Hd}$. We stress that similar
results are obtained by changing the values above by $\pm 30\gev$,
with similar comments applying to \eqref{S4} and \eqref{S5} below.

\vspace{0.2cm}
{\bf Scenario 4:}

\noindent
This is our favourite scenario in which the most interesting departures from the 
SM and MFV can be obtained and the problems addressed by us before can
be solved. In this scenario
\be\label{S4}
m_{H1}\approx m_{H2}= 500\gev\,,  \qquad     m_{H3}= 1000 \gev\,,
\ee
\be\label{S4a}
  \frac{1}{\sqrt{2}} \le s_{12}^d \le 0.99\,, \qquad 
             5\cdot 10^{-5}\le   s_{23}^d \le  2 \cdot 10^{-4}\,,
\qquad       4\cdot 10^{-2}\le  s_{13}^d \le 0.6
\ee
and the phase $\delta^d_{13}$ is arbitrary. 
The hierarchical structure of the CKM matrix
\be\label{CKMH}
s_{13}\ll s_{23}\ll s_{12}\,, \qquad (\text{CKM})
\ee
is changed in this scenario to
\be
s^d_{23}\ll s^d_{13}\le s^d_{12}\,, \qquad (V_{Hd})
\ee
so that $V_{Hd}$ looks as follows: 
\be\addtolength{\arraycolsep}{3pt}
V_{Hd} =  \begin{pmatrix}
 c^d_{12} & s^d_{12} & s^d_{13}e^{-i\delta^d_{13}}\\
-  s^d_{12} &  c^d_{12} &  s^d_{23}\\
- c^d_{12}  s^d_{13}e^{i\delta^d_{13}} & -  s^d_{12}  
s^d_{13}e^{i\delta^d_{13}} &1 \end{pmatrix}.
\ee
The very different structure of $V_{Hd}$
when compared with $V_{\rm CKM}$,
with a large complex phase in the $(V_{Hd})_{32}$ element assures large 
CP-violating effects in the $B^0_s-\bar B^0_s$ system without any problem with
$\Delta M_K$ as the first two mirror fermion masses are very close to each 
other. Furthermore $\Delta M_s$ can be smaller than its SM value in this scenario,
and interesting effects in the $B^0_d-\bar B^0_d$ system are also found.

\vspace{0.2cm}
{\bf Scenario 5:}

\noindent
In all the previous scenarios we will choose the first solution for
 the angle $\gamma$ from tree level decays as given in
 (\ref{eq:gamma}) below so 
 that only small departures from the SM in the $B^0_d-\bar B^0_d$ system 
 will be consistent with the data. Here we will  assume
 the second solution for $\gamma$ in (\ref{eq:gamma}) in contradiction 
 with the SM and MFV. We will then ask whether the presence of 
 new flavour violating interactions can still bring the theory to agree with 
 all available data, in particular with the asymmetry $S_{\psi K_S}$. 

 It turns out that for a particular choice of the parameters of the LHT 
 model, consistency with all existing data can be obtained, although 
 this scenario appears to be less likely than Scenario 4.

In this scenario
\be\label{S5}
m_{H1}=500\gev, \qquad m_{H2}= 450\gev,  \qquad     m_{H3}= 1000 \gev,
\ee
\be\label{S5a}
  5\cdot 10^{-5}\le s_{12}^d \le 0.015, \qquad 
             2\cdot 10^{-2}\le   s_{23}^d \le  4 \cdot 10^{-2},
\qquad       0.2 \le  s_{13}^d \le 0.5
\ee
and the phase $\delta^d_{13}$ arbitrary. 
We thus have an inverted hierarchy relative to the CKM one in (\ref{CKMH})
but also different from the one in scenario 4:
\be
s^d_{12}\le s^d_{23}\ll s^d_{13}, \qquad (V_{Hd}).
\ee
$V_{Hd}$ looks now as follows: 
\be\addtolength{\arraycolsep}{3pt}
V_{Hd} =  \begin{pmatrix}
 c^d_{13} & s^d_{12} c^d_{13} & s^d_{13}e^{-i\delta^d_{13}}\\
-  s^d_{12}  &  c^d_{12} &  s^d_{23} c^d_{13}\\
- s^d_{13}e^{i\delta^d_{13}} & -  s^d_{23}  
& c^d_{13} \end{pmatrix}\approx \begin{pmatrix} c^d_{13}& 0 &s^d_{13}e^{-i\delta^d_{13}}
   \\ 0&1&0 \\- s^d_{13}e^{i\delta^d_{13}} &0&c^d_{13} \end{pmatrix}.
\ee
The very different structure of $V_{Hd}$
when compared with $V_{\rm CKM}$,
allows to make this scenario compatible with the data.
The price one has to pay are tiny new physics effects in the $B_s^0-\bar B_s^0$ 
system.

\newsection{Numerical Analysis }
\label{sec:beyond}
\subsection{Preliminaries}
\label{subsec:5.1}

In our numerical analysis we will set $|V_{us}|$, $|V_{cb}|$ and
$|V_{ub}|$ to their central values measured in tree level
decays~\cite{BBpage,CKM2005} and collected in
Table~\ref{tab:input}.

\begin{table}[ht]
\renewcommand{\arraystretch}{1}\setlength{\arraycolsep}{1pt}
\center{\begin{tabular}{|l|l|}
\hline
{\small $G_F=1.16637\cdot 10^{-5} \gev^{-2}$} & {\small$\Delta M_K= 3.483(6)\cdot 10^{-15}\gev$} \\
{\small$\mw= 80.425(38)\gev$} & {\small$\Delta M_d=0.507(4)/ \rm{ps}$\hfill\cite{BBpage}} \\\cline{2-2}
{\small$\alpha=1/127.9$} &{\small $\Delta M_s = 17.4(4)/\text{ps}$\hfill\cite{CDFnew,D0}} \\\cline{2-2}
{\small$\sin^2 \theta_W=0.23120(15)$\qquad\hfill\cite{PDG}} & {\small
  $F_K\sqrt{\hat B_K}= 143(7)\mev$\qquad\hfill\cite{Hashimoto,PDG}}\\\hline
{\small$|V_{ub}|=0.00423(35)$} &  {\small $F_D\sqrt{\hat B_D}= 202(39)\mev$\hfill\cite{Okamoto:2005zg}}\\\cline{2-2}
{\small $\vcb = 0.0416(7)$\hfill\cite{BBpage}} & {\small$F_{B_d} \sqrt{\hat B_{B_d}}= 214(38)\mev$} \\\cline{1-1}
{\small$\lambda=|V_{us}|=0.225(1)$ \hfill\cite{CKM2005}} & {\small$F_{B_s} \sqrt{\hat B_{B_s}}= 262(35)\mev$\;\;\hfill\cite{Hashimoto}} \\\hline
 {\small$|V_{ts}|=0.0409(9)$ \hfill\cite{UTFIT}} & {\small$\eta_1=1.32(32)$\hfill\cite{eta1}} \\\hline
{\small$m_{K^0}= 497.65(2)\mev$} & {\small$\eta_3=0.47(5)$\hfill\cite{eta3}}\\\cline{2-2}
{\small$m_{D^0}=  1.8645(4)\gev$} &{\small$\eta_2=0.57(1)$} \\
{\small$m_{B_d}= 5.2794(5)\gev$} & {\small$\eta_B=0.55(1)$\hfill\cite{eta2B}}\\\cline{2-2}
{\small$m_{B_s}= 5.370(2)\gev$} & {\small$\mcb= 1.30(5)\gev$} \\
{\small $|\varepsilon_K|=2.284(14)\cdot 10^{-3}$ \hfill\cite{PDG}} &{\small$\mtb= 163.8(32)\gev$} \\
\cline{1-1}
{\small $S_{\psi K_S}=0.687(32)$ \hfill\cite{BBpage}} & \\
\hline
\end{tabular}  }
\caption {\textit{Values of the experimental and theoretical
    quantities used as input parameters.}}
\label{tab:input}
\renewcommand{\arraystretch}{1.0}
\end{table}

As the fourth parameter we will choose the angle $\gamma$ in the
standard UT that to an excellent approximation equals the
phase $\delta_\text{CKM}$ in the CKM matrix. The angle $\gamma$ has been 
extracted
from $B \rightarrow D^{(*)} K$ decays without the influence of 
new physics with the
result~\cite{UTFIT}
\be
\gamma=(71 \pm 16)^\circ\,, \qquad \gamma=-(109 \pm 16)^\circ\,.
\label{eq:gamma}
\ee
Only the first solution agrees with the SM analysis of the UT
but as we go beyond the SM in the present paper we will
investigate in Scenario 5 whether the second solution could be consistent
with the data within the LHT model. The error in the first
solution is sufficiently large to allow for significant
contributions from new physics.

For the non-perturbative parameters entering the analysis of
particle-antiparticle mixing we choose and collect in Table~\ref{tab:input}
their lattice averages given in~\cite{Hashimoto}, which combine unquenched
results obtained with different lattice actions.

In order to simplify our numerical analysis we will set all non-perturbative 
parameters to their central values and instead we will allow $\Delta M_K$, 
$\varepsilon_K$, $\Delta M_d$, $\Delta M_s$ and $S_{\psi K_S}$ to differ from 
their experimental values by $\pm 50\%$, $\pm 40\%$, $\pm 40\%$, $\pm 40\%$ 
and $\pm 8\%$, respectively. 
In the case of $\Delta M_s/\Delta M_d$ we will choose $\pm 20\%$, as the error
on the relevant parameter $\xi$ is smaller than in the case of $\Delta M_d$
and $\Delta M_s$ separately.
This could appear rather conservative, but we 
do not want to miss any interesting effects by choosing too optimistic 
non-perturbative uncertainties.
In Scenarios $3-5$, then, the parameters $f$ and $x_L$ will be fixed to
$f=1000\gev$ and $x_L=0.5$ in accordance with electroweak precision tests~\cite{mH}.

\subsection{Scenario 1}
\label{subsec:5.2}

Let us consider first the case of totally degenerate mirror
fermions. In this case the odd contributions vanish due to the GIM
mechanism~\cite{GIM}, the only new particle contributing is $T_+$
and the LHT model in this limit belongs to the class of MFV
models. As only the T-even sector contributes, the new
contributions to particle-antiparticle mixing and $B \rightarrow
X_s \gamma$ are entirely dependent on only two parameters 
\be
x_L\,,\qquad f\,. \ee

Moreover, all the dependence on new physics contributions is
encoded in the function $S_t$ in \eqref{3.4} in the case of
particle-antiparticle mixing and the functions  $T^{\rm even}_{D'}$
and  $T^{\rm even}_{E'}$
in \eqref{eq:TDeven} and \eqref{eq:TEeven} in the case of the $B\to X_s\gamma$ decay. 

\begin{figure}
\begin{minipage}{7.9cm}
\center{\epsfig{file=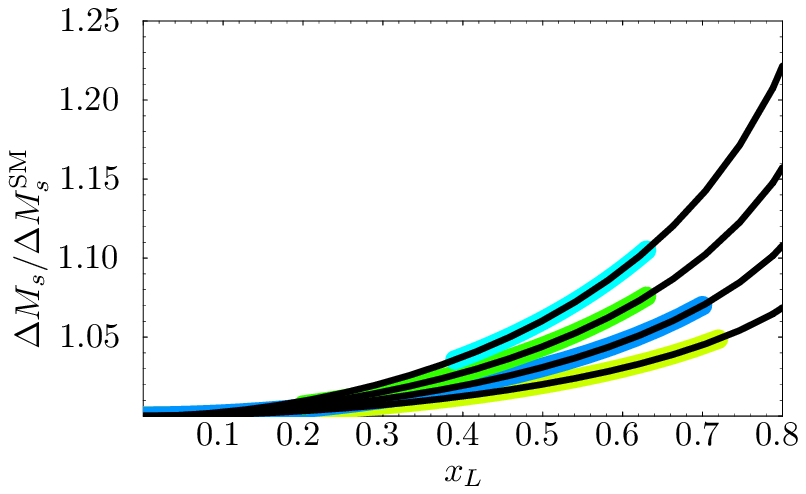,scale=0.9}}
\end{minipage}\hfill
\begin{minipage}{7.9cm}
\center{\epsfig{file=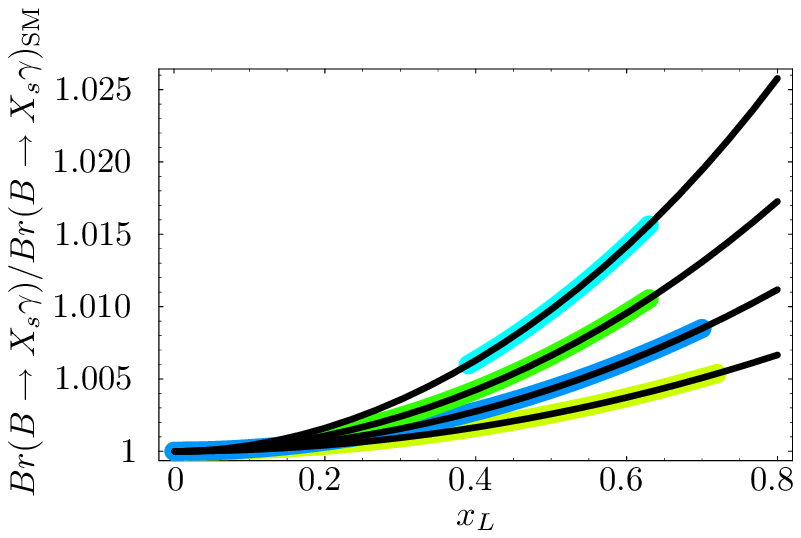,scale=0.9}}
\end{minipage}
\caption{\textit{$\Delta M_s/(\Delta M_s)_{\rm SM}$ and  $Br(B\to X_s\gamma)/Br(B\to X_s\gamma)_\text{SM}$ in Scenario 1 as  functions of
$x_L$ for values of $f=1, 1.2, 1.5,$ and $2$ TeV from top to
bottom. The bands underlying the curves show the allowed ranges after
applying electroweak precision constraints \cite{mH}.}}
\label{fig:Scen1.1}
\end{figure}

It should be emphasized that in this scenario  the ``problems''
listed in (\ref{P1}) cannot be solved as it is a MFV scenario.
Moreover, $\Delta M_s\ge (\Delta M_s)_{\rm SM}$, which is not
favoured by the CDF measurement.
Also $\Delta M_d\ge (\Delta M_d)_{\rm SM}$ in this scenario.
  Therefore in
Fig.~\ref{fig:Scen1.1} we
show the ratio $\Delta M_s/(\Delta M_s)_{\rm SM}$ and the corresponding 
ratio for $Br(B\to X_s\gamma)$ as  functions of
$x_L$ for various values of $f$.

We find that the maximal relative enhancements with respect to the SM are $13\%$ for $\Delta M_s$ and about $1.5\%$ for  $Br(B\to X_s\gamma)$. In view of
large theoretical errors in the evaluation of $\Delta M_{d,s}$, however, this
scenario cannot be tested at present. Similarly to $Br(B\to
X_s\gamma)$, the new physics effects in $B\to X_d\gamma$ and the
corresponding two CP asymmetries are very small.

\subsection{Scenario 2}
\label{subsec:5.3}
 At first sight one could think that this is another version of the MFV 
    scenario just discussed, but this is not the case. 
The point is that breaking the
    degeneracy of mirror fermion masses introduces a new source of 
 flavour violation that has nothing to do with the top Yukawa couplings. 
Only if accidentally the contributions proportional to $\xi^{(q)}_3$  
dominates the new physics contributions, one would again end up with a 
scenario that effectively looks like MFV. However, as the mirror spectrum is
generally different from the quark spectrum and not as hierarchical as the 
latter one, the terms involving $\xi^{(q)}_2$ in the formulae
\eqref{3.17}-\eqref{3.21}, \eqref{eq:TDodd} and \eqref{eq:TEodd}
 cannot be neglected although 
this can be done in the T-even contributions. As the phases in 
$\lambda_c^{(q)}$ are different from the ones in $\lambda_t^{(q)}$, that dominate the 
SM contributions, even in this simple scenario the MFV relations in \eqref{eq:KdsMFV} 
will be violated.

The new contributions to particle-antiparticle mixing and $B
\rightarrow X_s \gamma$ are in this scenario entirely dependent on only five
parameters
\be
x_L\,,\qquad f\,,\qquad 
m_{H1}\,,\qquad m_{H2}\,, \qquad m_{H3}
\ee
in addition to $m_t$ and on the CKM parameters that we set to the central 
values obtained from tree level decays. 

\begin{figure}
\begin{minipage}{7.9cm}
\center{\epsfig{file=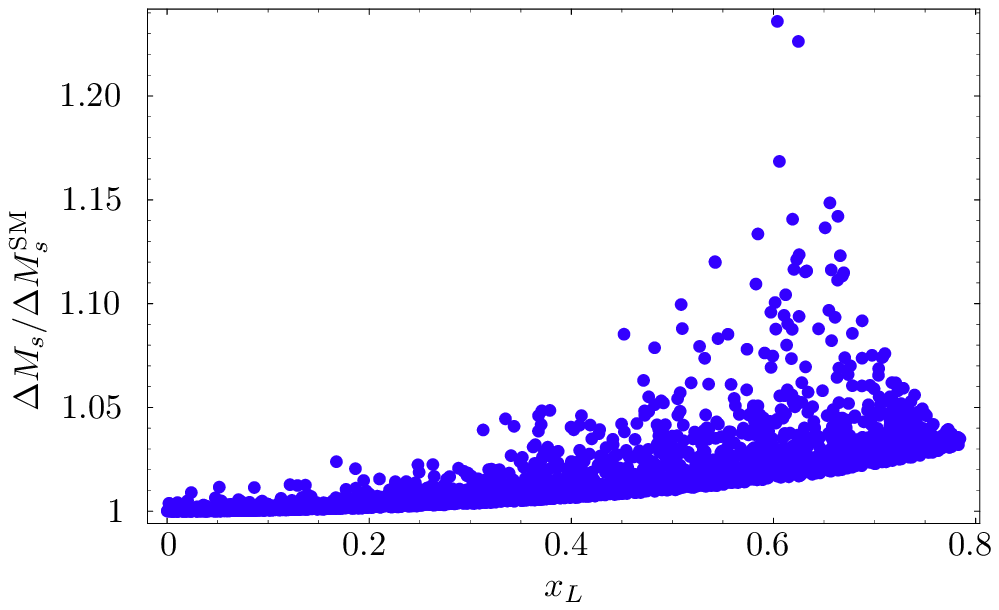,scale=0.75}}
\end{minipage}\hfill
\begin{minipage}{7.9cm}
\center{\epsfig{file=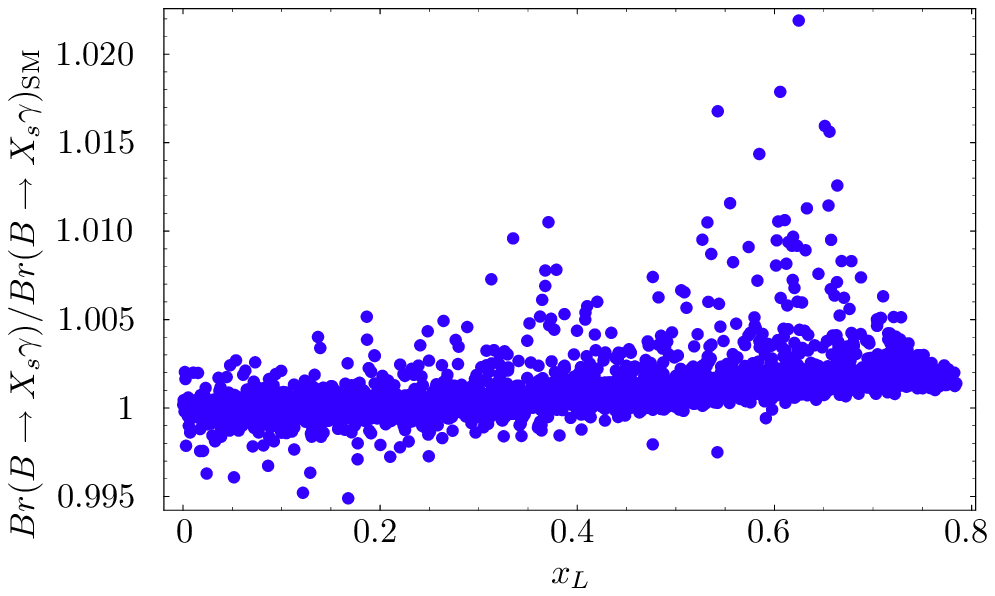,scale=0.75}}
\end{minipage}
\caption{\textit{$\Delta M_s/(\Delta M_s)_{\rm SM}$ and  $Br(B\to X_s\gamma)/Br(B\to X_s\gamma)_\text{SM}$ as  functions of
$x_L$ in Scenario 2.}}
\label{fig:Scen2.1}
\end{figure}

Our numerical analysis shows that also in this scenario none of the 
problems listed in Section~\ref{sec:goals} can be solved. Still 
the new physics effects are larger than in the Scenario 1 just 
discussed. This is shown in Fig.~\ref{fig:Scen2.1} which corresponds
to Fig.~\ref{fig:Scen1.1}.
 
We have evaluated the relative deviations of $\Delta M_s$ and $Br(B\to
X_s\gamma)$ from their SM values for mirror fermion masses in the 
range of $300$\dots$3000$ GeV and for combinations of $x_L$ and $f$
allowed by precision electroweak constraints. One finds that the new
contributions by the mirror fermions additionally enhance the ratios
$\Delta M_q/(\Delta M_q)_{\rm SM}$ for all and the ratio $Br(B\to
X_s\gamma)/Br(B\to X_s\gamma)_\text{SM}$ for most choices of the
mirror spectrum. The maximal deviations are $20\%$ and $2 \%$
respectively. The new physics effects in $Br(B\to X_d\gamma)$ and in
$A_\text{CP}(B\to X_{s,d}\gamma)$ are very small.  The only interesting constraint for this choice of
$V_{Hd}$ is the bound on the mass splitting between the first two
mirror quark generations coming from  $\Delta M_K$ and $\varepsilon_K$ as discussed in~\cite{Hubisz}.

\subsection{Scenario 3}
\label{subsec:7.3}

In this scenario, relative to the previous scenarios, there is the first
hope that our problems could be solved as the matrix $V_{Hd}$ now differs 
from the CKM matrix. In particular 
\be S_{\psi K_S}=\sin (2\beta_{\rm true}+2\varphi_{B_d})\,,
\ee 
where $\beta_{\rm true}=25.8^\circ$ gives $S_{\psi K_s}=0.78$ for
$\varphi_{B_d}=0$. Thus in order to fit the experiment we need a small
negative phase $\varphi_{B_d}$. It turns out that 
\begin{itemize}
\item
In this scenario $\varphi_{B_d}$ is  consistent with all existing
constraints in the range of $[-45^\circ,45^\circ]$  and 
it is possible
to obtain agreement with the experimental value of $S_{\psi K_S}$.
\item
Interestingly, also in this scenario, $\Delta M_s\ge(\Delta M_s)_{\rm
  SM}$ with maximal deviations from the SM around $15\%$. $Br(B\to
X_s\gamma)$ can be enhanced by at most $3\%$ and suppressed by $1\%$. Similarly to $Br(B\to
X_s\gamma)$, the new physics effects in $B\to X_d\gamma$ and the
corresponding two CP asymmetries are very small.
\item
CP-violating effects in the $B_s^0-\bar B_s^0$ system are small since $\varphi_{B_s}$ is in the ballpark of 
$\pm 2^\circ$.
\end{itemize}

\subsection{Scenario 4}
\label{subsec:6.6}

We have seen that except for the solution of the ``$R_b-\sin
2\beta$'' problem in Scenario 3 none of the goals  on our list could  be reached in 
the three scenarios considered so far and it is time to modify 
the mirror fermion spectrum and the structure of the matrix $V_{Hd}$ in 
order to make any progress. In particular
the CP-violating effects in the
$B^0_s-\bar B^0_s$ system remained small. This is easy to understand. If 
the matrix $V_{Hd}$ has a hierarchical structure that is similar to the 
one of the CKM matrix, the phases of $\xi^{(s)}_2$ and $\xi^{(s)}_3$ that are 
relevant for  CP violation in the $B^0_s$ system will remain small. 
In order to obtain large CP-violating effects in this system we have to
increase the phases of these two CKM-like factors. While doing this we 
have to make sure that the known CP-violating effects in the $B^0_d$ and $K^0$ 
systems are still consistent with the data. The case of $B^0_d$ is not 
very problematic as the CP-violating effects are large anyway, 
but due to 
small experimental values of $\varepsilon_K$ and $\Delta M_K$ only 
for a particular structure of $V_{Hd}$ we can reach goal 3 without 
disagreeing with the data on these two observables. 

\begin{figure}
\begin{minipage}{7.9cm}
\center{\epsfig{file=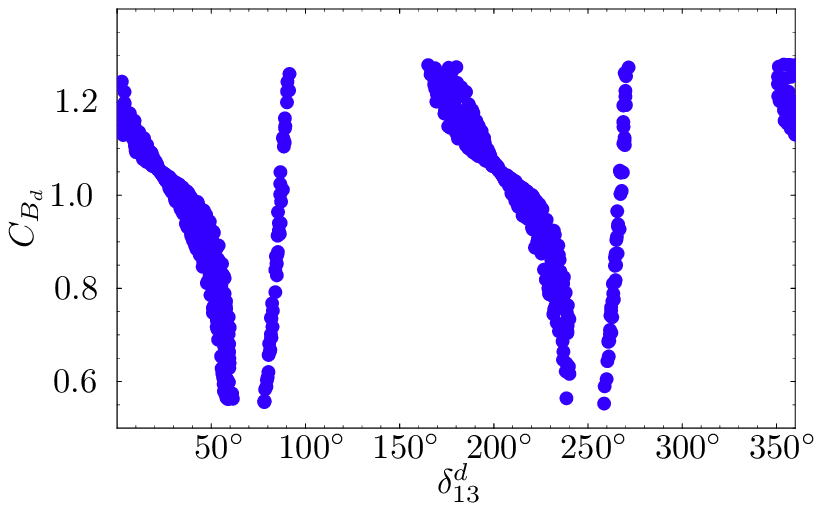,scale=0.9}}
\end{minipage}\hfill
\begin{minipage}{7.9cm}
\center{\epsfig{file=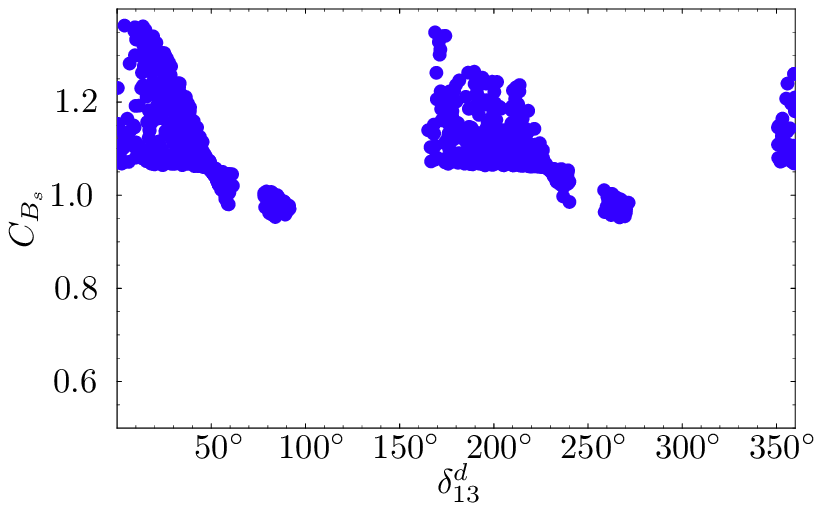,scale=0.9}}
\end{minipage}
\caption{\textit{$C_{B_d}$ and $C_{B_s}$ as functions 
of the new phase $\delta_{13}^d$ in Scenario 4.}}
\label{Cdelta}
\end{figure}

\begin{figure}
\center{\epsfig{file=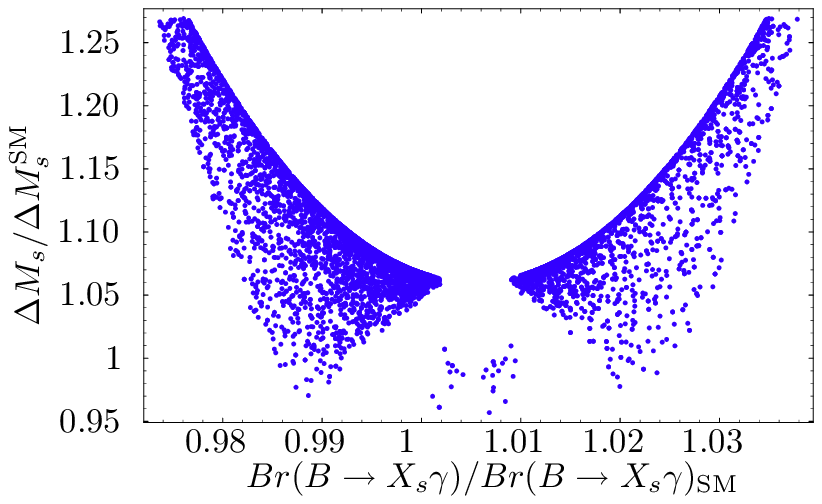,scale=1}}
\caption{ \textit{Correlation between $C_{B_s}$ and $Br(B\to X_s\gamma)$ in Scenario 4.}}
\label{Bsg}
\end{figure}

\begin{figure}
\center{\epsfig{file=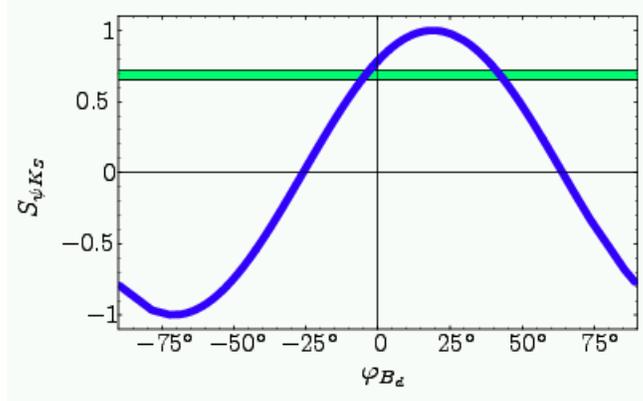,scale=0.6}}
\caption{\textit{$S_{\psi K_S}$ as 
a function of $\varphi_{B_d}$ in Scenario 4.}}
\label{Sphid}
\end{figure}

\begin{figure}
\begin{minipage}{7.9cm}
\center{\epsfig{file=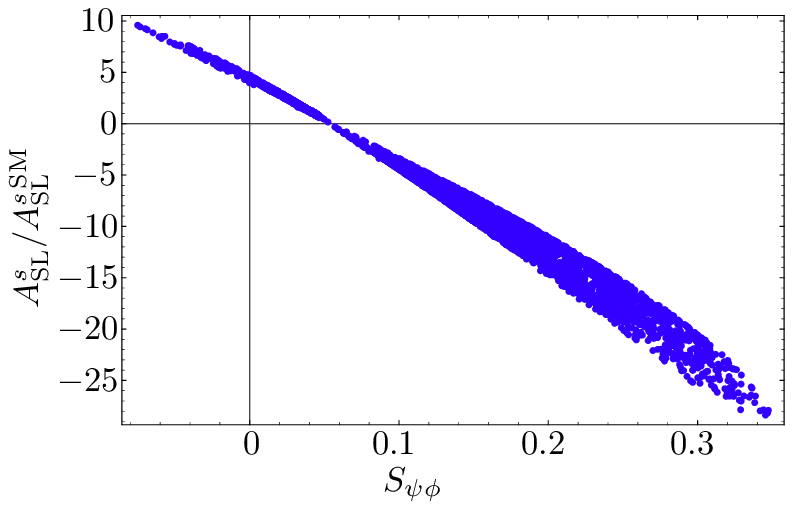,scale=0.9}}
\end{minipage}\hfill
\begin{minipage}{7.9cm}
\center{\epsfig{file=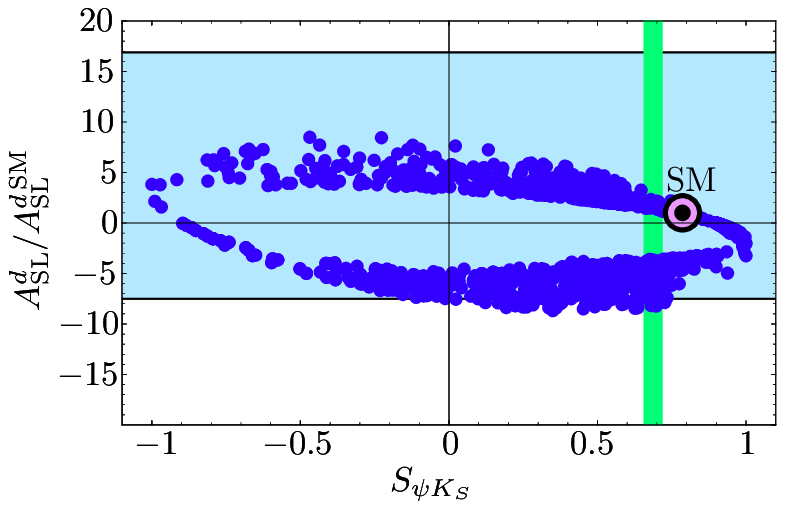,scale=0.9}}
\end{minipage}
\caption{\textit{$A_{\rm SL}^s$ and $A_\text{SL}^d$ as functions of
  $S_{\psi\phi}$ and $S_{\psi K_S}$, respectively, in Scenario 4. The
  shaded areas represent the experimental data.}}
\label{ASL}
\end{figure}

\begin{figure}
\center{\epsfig{file= 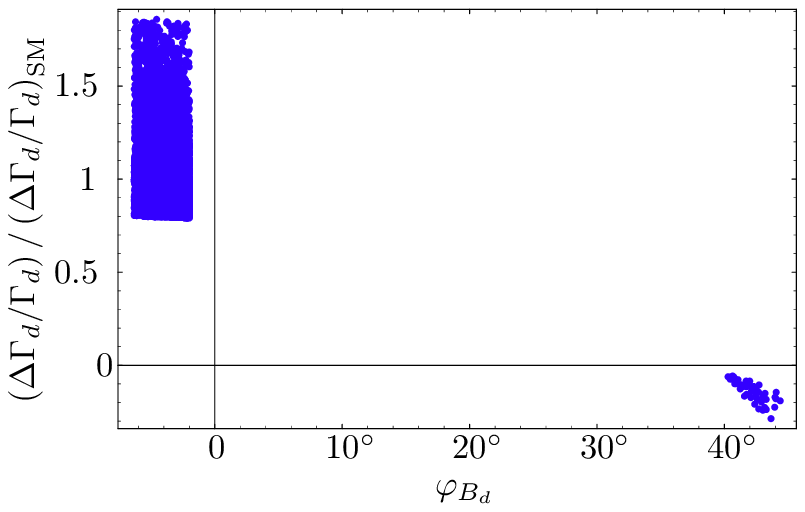,scale=1}}
\caption{\textit{$\Delta\Gamma_d/\Gamma_d$ as a function of 
$\varphi_{B_d}$ in Scenario 4.}}
\label{DGsd}
\end{figure}

By inspecting the matrix $V_{Hd}$ in (\ref{2.12a}) we conclude that 
the mirror fermions in the first two generations have to be almost 
degenerate in mass in order to satisfy the $\Delta M_K$ and $\varepsilon_K$ 
constraints 
and simultaneously the mixing parameters in the $V_{Hd}$ matrix must
be in the ranges given in (\ref{S4a}). This simple procedure turns out
to be successful: all our goals can be reached. The most interesting 
results in this scenario are collected in  Figs.~\ref{Cdelta}--\ref{DGsd}.

In Fig.~\ref{Cdelta} we show the ratios $C_{B_d}$ and $C_{B_s}$ as functions 
of the new phase $\delta_{13}^d$. We observe that only certain ranges of 
$\delta_{13}^d$ are allowed. This follows from the experimental constraint 
on $S_{\psi K_S}$. We also observe that while it is easy to obtain values 
of $C_{B_d}$ below unity, this is much harder in the case of $C_{B_s}$. Yet 
a suppression of $\Delta M_s$ relative to $(\Delta M_s)_{\rm SM}$ by $5-10\%$ 
is possible in this scenario, which should be sufficient to obtain an 
agreement with experiment if necessary.
We observe that the suppression below unity is more likely 
in the case of $C_{B_d}$. The ratio 
$C_{B_d}/C_{B_s}$ can deviate from unity even by $(30-40)\%$ so that 
a relevant violation of the MFV relation between $\Delta M_s/\Delta M_d$ 
and $|V_{ts}/V_{td}|$ as seen in  (\ref{3.35}) is possible.

In Fig.~\ref{Bsg} we show the correlation between $C_{B_s}$ and 
$Br(B\to X_s\gamma)$ normalized to its central SM value. The main message
from this plot is that $Br(B\to X_s\gamma)$ is changed by at most $\pm 4\%$
which is welcomed as the SM agrees well with the data. It will 
be very difficult to distinguish LHT from the SM in this case. 
New physics effects in $Br(B\to X_d\gamma)$ and $A_\text{CP}(B\to
X_{s,d}\gamma)$ are small. 

More interesting effects are found in the CP-violating observables related 
to $B^0_d-\bar B^0_d$ mixing and in particular to $B^0_s-\bar B^0_s$ mixing.
As already seen in Fig.~\ref{Cdelta}, this scenario is consistent with the 
data on $S_{\psi K_S}$ in spite of a large value of $R_b$. In order to 
illustrate this explicitly, we show in Fig.~\ref{Sphid} $S_{\psi K_S}$ as 
a function of $\varphi_{B_d}$. To this end we have removed the corresponding 
experimental constraint but kept the remaining ones. For 
$\varphi_{B_d}\approx -3^\circ$ and $\varphi_{B_d}\approx 43^\circ$
 agreement with experiment can be obtained. We will see below that 
the second solution although not ruled out is not favoured  by the data on 
$A_{\rm SL}^d$. 
At present, therefore, the cosine measurement
$\cos(2\beta+2\varphi_{B_d})=1.69 \pm 0.67$ \cite{BBpage} represents the
strongest constraint in disfavouring the solution
$\varphi_{B_d}=43^\circ$.

While this result is clearly interesting, an even more impressive result is 
shown in the left panel of Fig.~\ref{ASL}, where we plot $A_{\rm SL}^s$ normalized to its 
SM central value versus $S_{\psi\phi}$. Comparing this plot with the 
corresponding 
plot in \cite{Ligeti}, where the correlation between
$A_{\rm SL}^s$ and $S_{\psi\phi}$ has been pointed out, we observe that in 
a specific model like the one considered here, the correlation in question is 
much stronger than  in a model independent approach considered in 
that paper. This plot shows that $S_{\psi\phi}$ can be as large as  
$+0.30$ and the absolute value of the asymmetry $A_{\rm SL}^s$  can be 
enhanced by a factor of $10-20$ relative to the SM value. While both 
asymmetries can have both signs, $S_{\psi\phi}>0$ and $A_{\rm SL}^s <0$ seem to 
be more likely, which implies the preference for a {\it negative} phase
$\varphi_{B_s}$. The present data are not yet conclusive but the analysis 
in \cite{Nir} indicates that this sign is also favoured by the data on 
$A_{\rm SL}^s$.

In the right panel of Fig.~\ref{ASL} we show  $A_{\rm SL}^d$ normalized to its 
SM value versus $S_{\psi K_S}$. As the latter asymmetry is already well 
measured, the new physics effects are much more constrained
 than in the case 
of $A_{\rm SL}^s$. Still an enhancement by a factor of 3 for the case of 
$\varphi_{B_d}\approx -3^\circ$ is possible. On the other hand as seen in 
Fig.~\ref{ASL} for the $\varphi_{B_d}\approx 43^\circ$ solution the asymmetry 
in question changes sign relatively to the SM value and its magnitude can be 
enhanced by a factor of seven, which could  soon be ruled out with improved 
data.

Finally, in Fig.~\ref{DGsd} we show $\Delta\Gamma_d/\Gamma_d$ versus 
$\varphi_{B_d}$. The experimental error in \eqref{3.43b} is so large
that nothing conclusive can be said at present. The future improved
data could help to distinguish between the two solutions for
$\varphi_{B_d}$. In the case of $\Delta \Gamma_s/\Gamma_s$, the SM 
value, that is below the experimental data, is further suppressed 
for $\varphi_{B_s}\not=0$, but even for $\varphi_{B_s}=-8^\circ$ corresponding to 
$S_{\psi\phi}=0.30$, this suppression amounts to  $5\%$. 
Improved data and the theory for $\Delta\Gamma_s$ will tell us 
how large the phase $\varphi_{B_s}$ and the asymmetry $S_{\psi\phi}$ 
can be.

\subsection{Scenario 5}
\label{subsec:6.7}

\begin{figure}
\center{\epsfig{file=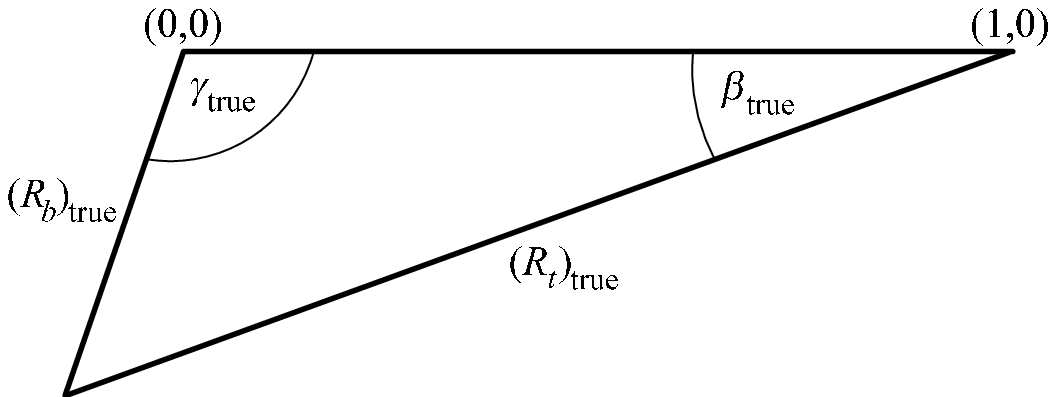,scale=1}}
\caption{\textit{``Upside-down'' unitarity triangle in Scenario 5.}}
\label{fig:upside-down}
\end{figure}

Using the central value for $\gamma$ in the second solution in (\ref{eq:gamma})
and the central value of $|V_{ub}|$  in Table~\ref{tab:input}
 we  find by means of (\ref{VUBG})
\be\label{crazy}
(R_t)_{\rm true}= 1.217\,, \qquad \beta_{\rm true}=-20.0^\circ\,,
\ee
and consequently within the SM approximately opposite signs for 
 $\varepsilon_K$ 
and $S_{\psi K_S}$ compared with the data: 
    \be\label{wrong1}
\varepsilon_K= -3.72\cdot10^{-3}\,e^{i\pi/4}\,, \qquad  S_{\psi K_S}= -0.643\,.
\ee
The corresponding unitarity triangle is shown in
Fig.~\ref{fig:upside-down}.

In order to obtain agreement with the data we need positive new physics
contributions in both cases that are in magnitude by a factor of three and
 two, respectively,
larger than the SM contribution. 
On the other hand  $\Delta M_d$ turns out to be too large
\be\label{wrong2}
\Delta M_d=0.904/{\rm ps}\,.
\ee
The question then arises whether one could still modify all these values with
the help of mirror fermions. As now a very large {\it positive} 
phase $\varphi_{B_d}$ is 
required to fit the experimental value of $S_{\psi K_S}$, it will be 
interesting
to see  how $\Delta \Gamma_q$  given in (\ref{3.44a}) is modified 
relatively to the 
SM value.

\begin{figure}
\center{\epsfig{file=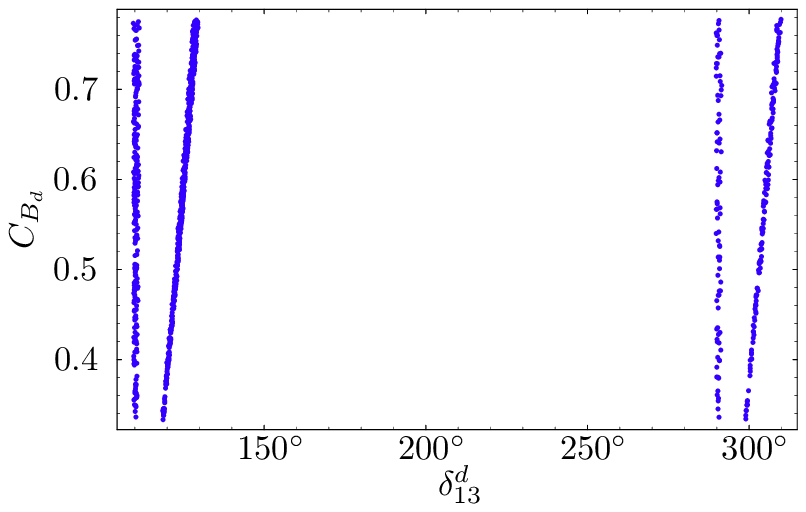,scale=1}}
\caption{\textit{$C_{B_d}$ as a function of $\delta^d_{13}$ in Scenario 5.}}
\label{5Cddel}
\end{figure}

\begin{figure}
\center{\epsfig{file=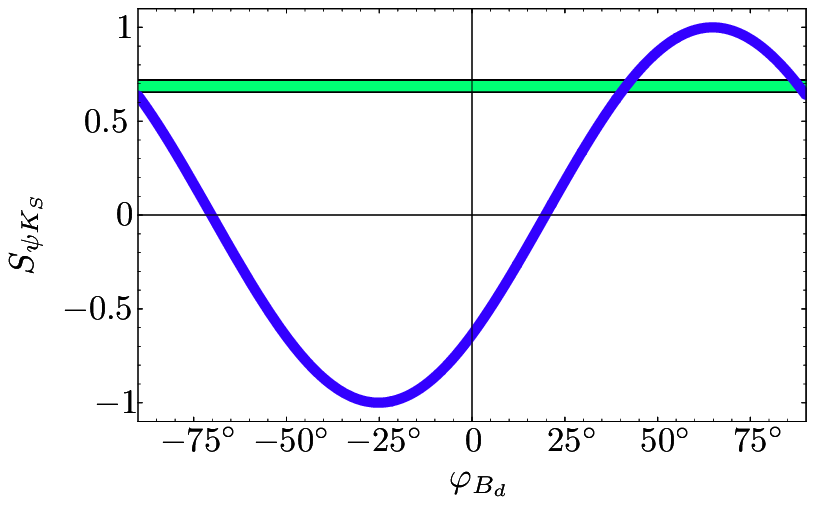,scale=1}}
\caption{\textit{$S_{\psi K_S}$ as a function of $\varphi_{B_d}$  in Scenario 5.}}
\label{5psiKS}
\end{figure}

\begin{figure}
\center{\epsfig{file=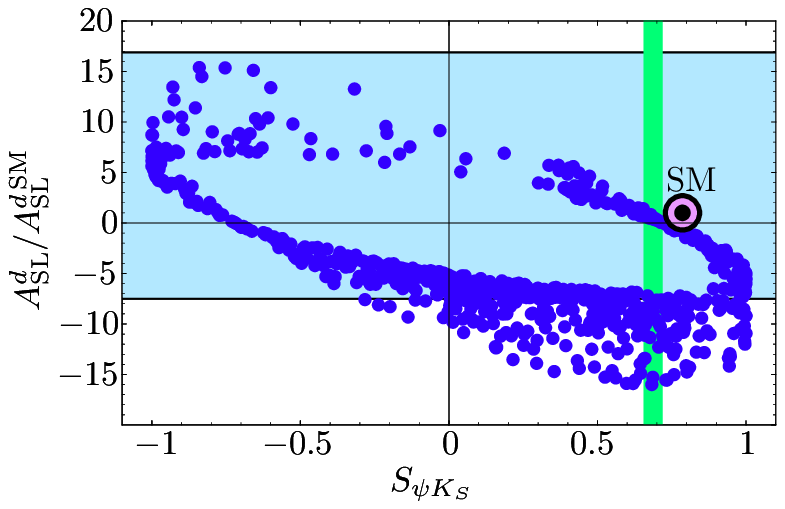,scale=1}}
\caption{\textit{$A_{\rm SL}^d$  as a function of 
$S_{\psi K_S}$  in Scenario 5. The shaded area represents the
experimental data.}}
\label{5ASLd}
\end{figure}

\begin{figure}
\center{\epsfig{file=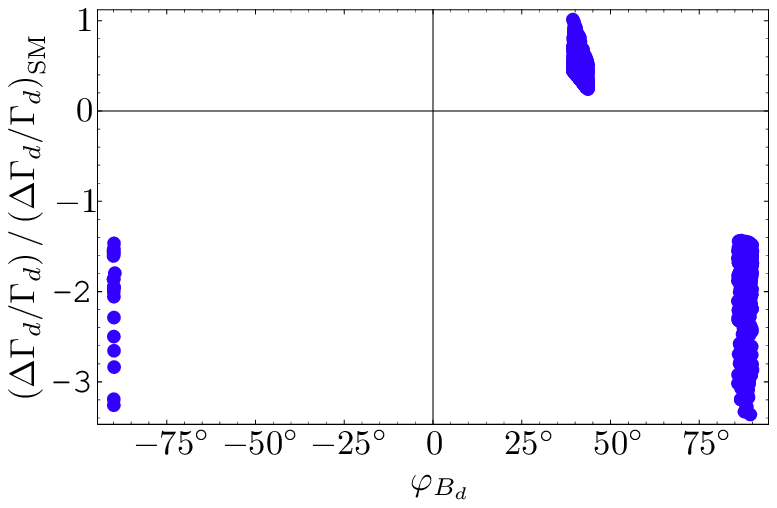,scale=1}}
\caption{\textit{$\Delta\Gamma_d/\Gamma_d$ as a function of 
$\varphi_{B_d}$  in Scenario 5.}}
\label{5Gd}
\end{figure}

The most interesting results in this scenario, related to the 
$B_d^0-\bar B_d^0$ system, are shown in 
Figs.~\ref{5Cddel}--\ref{5Gd}.

As already stated in the previous section, in this scenario the CP-violating 
effects in the $B^0_s-\bar B^0_s$ system are very small. It also turns out 
that in this scenario $\Delta M_s$ cannot be suppressed relative to the SM. 
On the other hand as clearly seen in Fig.~\ref{5Cddel}, where we plot 
$C_{B_d}$ versus $\delta^d_{13}$, $C_{B_d}$ can be suppressed below unity
bringing $\Delta M_d$ to agree with experiment.

The CP-violating new physics effects in the 
$B^0_d-\bar B^0_d$ system are spectacular in this scenario 
because the phase $\varphi_{B_d}$ 
must take a value in the ballpark of $42^\circ$ or $88^\circ$  
in order to fit 
the experimental value of $S_{\psi K_S}$. We show this in Fig.~\ref{5psiKS}.

In turn the large value of $\varphi_{B_d}$ has a large impact on $A_{\rm SL}^d$ 
and $\Delta\Gamma_d$. This study requires, however, some care. 
The point is that with the value $\gamma=-109^\circ$, also the values 
in (\ref{eq:r2}) and (\ref{eq:r1}) change. We find now
\begin{gather}
\text{Re}\left
  (\frac{\Gamma_{12}^d}{M_{12}^d} \right) = -(3.4 \pm 1.0)\cdot10^{-3}\,,\qquad
\text{Re}\left
  (\frac{\Gamma_{12}^s}{M_{12}^s} \right) = -(2.6 \pm 1.0)\cdot10^{-3} \,,
\label{eq:r5}\\
\text{Im}\left
  (\frac{\Gamma_{12}^d}{M_{12}^d} \right) = +(3.8 \pm 0.8)\cdot
10^{-4}\,,\qquad \text{Im}\left
  (\frac{\Gamma_{12}^s}{M_{12}^s} \right) = -(3.2 \pm 0.6)\cdot 10^{-5}\,,
\label{eq:r6}
\end{gather}

In Fig.~\ref{5ASLd} we show 
$A_{\rm SL}^d$  as a function of 
$S_{\psi K_S}$ normalized to the central SM value in (\ref{eq:r3}). 
The solution with large values of $A_{\rm SL}^d$ in 
Fig.~\ref{5ASLd} corresponds to $\varphi_{B_d}\approx 42^\circ$ and is
by an order of magnitude larger than the SM predictions and has 
the opposite sign. We would like to stress that, being $\beta =-20.0^\circ$ in this scenario, this is the solution
strongly favoured by the cosine measurement
$\cos(2\beta+2\varphi_{B_d})=1.69 \pm 0.67$ \cite{BBpage}.
Therefore, more accurate measurements of the semileptonic asymmetry
$A_{\rm SL}^d$ could soon rule out the $\varphi_{B_d}\approx 42^\circ$
solution and, when combined with the cosine measurement, the whole scenario
$5$.

In  Fig.~\ref{5Gd} we show
$\Delta\Gamma_d/\Gamma_d$ as a function of 
$\varphi_{B_d}$ normalized to the SM value in (\ref{eq:r4}). 
The large experimental 
errors do not allow to exclude any of these solutions at present.  

The new physics effects in $B\to X_{s,d}\gamma$ are very
small. However as $\gamma$ has been changed to $-109^\circ$ in the SM
contributions, the final results differ from the SM
expectations. $Br(B\to X_s\gamma)$ is suppressed by roughly  $4\%$ and
$Br(B\to X_d\gamma)$ enhanced by roughly a factor of
two. $A_\text{CP}(B\to X_d\gamma)$ is changed from $-10\%$ to $+5\%$
which could in principle be used to confirm or rule out this
scenario. On the other hand  $A_\text{CP}(B\to X_s\gamma)$ changes
sign but remains of the same size as in the SM.

\subsection{Comparison of various Scenarios}

The messages from this analysis are as follows:
\begin{itemize}
\item
Scenarios 1 and 2 are not very exciting and both are rather close 
to the SM expectations. In particular, they do not solve any of the 
problems listed in Section~\ref{sec:goals}.
\item
Scenario 3 is capable of solving the ``$R_b-\sin 2\beta$''  problem
but the problems of a too small $\Delta M_s$ in (\ref{P2}) and the 
smallness of 
CP-violating effects in the 
$B^0_s-\bar B^0_s$ system remain essentially unchanged.
\item
Scenario 4 appears to be the most interesting one as it offers 
solutions to all problems and predicts large CP-violating effects 
in the  $B^0_s-\bar B^0_s$ system. In particular, we find that $S_{\psi \phi}$ 
can be as large as $0.30$ and $A_{\rm SL}^s$ enhanced by more than an order 
of magnitude above the SM prediction. 
The plots in Fig.~\ref{ASL} demonstrate it in an impressive 
manner.
\item
Scenario 5 is also interesting as the presence of mirror fermions
allows for the agreement of the ``upside-down'' reference unitarity 
triangle, shown in Fig.~\ref{fig:upside-down}, with all existing data. 
We emphasize that in this scenario the asymmetry $A_\text{SL}^d$ has opposite
sign to the SM and, as seen in Fig.~\ref{5ASLd}, a more accurate measurement of
$A_\text{SL}^d$ could soon rule out this scenario.
\end{itemize}

\subsection{Determining the Parameters of the LHT Model}
\label{subsec:6.2}
The determination of the parameters of the LHT model is a very difficult
  experimental task as it would require first of all the discovery and the
  mass measurement of at least one
  heavy gauge boson, one heavy $T_\pm$ fermion and of three mirror fermions. 
The discussion of this issue is clearly
 beyond the scope of our paper which deals entirely with flavour physics.
 We will only indicate how the six parameters of the $V_{Hd}$
 matrix could be in principle determined through six FCNC processes, 
up to discrete ambiguities. 
As these parameters 
describe the deviations from the SM results, precise results obtained in the 
SM are required. From the present perspective, the mass differences 
$\Delta M_K$, $\Delta M_d$ and $\Delta M_s$ being still subjects to 
significant non-perturbative uncertainties, will not serve us in this decade
to achieve this goal. Among the observables related to particle-antiparticle 
mixing the following four stand out as being very useful
\be\label{best}
\frac{\Delta M_d}{\Delta M_s}, \qquad S_{\psi K_S}, \qquad S_{\psi\phi}, 
\qquad \varepsilon_K,
\ee
provided the accuracy on the parameters $\xi$ and in particular 
$\hat B_K$ will be further improved. Similarly, when the theoretical 
errors on $ {\rm Re}(\Gamma_{12}^q/M_{12}^q)^{\rm SM} $ decrease with 
time, the measurements of $A_{\rm SL}^q$ will determine the parameters
$C_{B_d}$ and $C_{B_s}$ as discussed in Section~\ref{3.8} and in \cite{BBGT},
provided 
$S_{\psi K_S}$ and $S_{\psi\phi}$ will differ significantly from the SM 
predictions. 
 
Additional information that can be used to determine the $V_{Hd}$ matrix 
will come one day from $Br(B_s\to \mu^+\mu^-)$ and $Br(B_d\to \mu^+\mu^-)$.
Their ratio does not depend on weak decay constants \cite{MFVrel} and is 
theoretically rather clean. Similar comments apply to 
$Br(B\to X_{s,d}\nu\bar\nu)$ and $Br(B\to X_{s,d}\ell^+\ell^-)$. 
Finally, at the beginning of the next decade 
the very clean decays $K^+\to\pi^+\nu\bar\nu$ and $K_L\to\pi^0\nu\bar\nu$ 
and useful decays $K_L\to \pi^0\ell^+\ell^-$ could provide decisive tests of 
the LHT model.
A detailed analysis of all these processes in the LHT model
  has been presented very recently in \cite{BBPRTUW}. The analytic expressions
  for the short distance functions $X$, $Y$ and $Z$ given there should
  allow to perform this program once the data on several  FCNC
  processes listed above will be available.

\newsection{\boldmath $D^0 - \bar D^0$ Mixing}
\label{sec:DD}

$D^0 - \bar D^0$ mixing in the SM has a very different structure
from $K^0 - \bar K^0$ and $B^0_q - \bar B^0_q$ mixings. Here the
quarks running in box diagrams are the down-type quarks implying that
the short distance part of $\Delta M_D$ is very strongly
suppressed in the SM by GIM. As a result of this structure,
$\Delta M_D$ in the SM is dominated by the long distance
contributions and unless new physics contributions are very large,
$\Delta M_D$ does not provide a useful constraint.

In the LHT model the T-even contributions to $\Delta M_D$ can be
neglected with the same argument as the SM contributions. Also
tree level effects, which appear due to the modified flavour
structure in the up-type quark sector, relatively to the SM, can
be neglected, as shown in \cite{Lee:2004me}. On the other hand, as
already analyzed in~\cite{Hubisz}, the mirror fermions could have
a significant impact on $\Delta M_D$.

The effective Hamiltonian for the mirror fermion
contribution to the $D^0 - \bar D^0$ system can be obtained from the $\Delta S
=2$ Hamiltonian in~(\ref{3.14}) with the following replacements:
\begin{gather}
\xi_i \rightarrow \xi_i^{(D)}={V_{Hu}^{ui}}^*\,V_{Hu}^{ci}\,,\qquad \sum_i
\xi_i^{(D)}=0\,,\label{eq:xiD}\\
(\bar s d)_{V-A}\,(\bar s d)_{V-A} \rightarrow (\bar c
u)_{V-A}\,(\bar c u)_{V-A}\,.\label{eq:QD}
\end{gather}
Therefore, in (\ref{3.17}) also the replacements \be F_K \rightarrow
F_D\,,\qquad \hat B_K \rightarrow \hat B_D\,,\qquad m_K
\rightarrow m_D\,, \label{eq:parD} \ee have to be made. The QCD
correction is approximately equal to $\eta_2$. Although the role
of up mirror fermions and down mirror fermions is now
interchanged, with down mirror fermions accompanied by $W_H^\pm$
and up mirror fermions accompanied by $Z_H$ and $A_H$ in box
diagrams, no change in~(\ref{3.18}), except for~(\ref{eq:xiD}),
has to be made, because of the equality of the masses of up and
down mirror fermions belonging to a given $SU(2)_L$ doublet.

The current experimental bound on $D^0-\bar D^0$ mixing is given by \cite{PDG}
\be\label{Dbound}
\Delta M_D = |m_{D^0_1}-m_{D^0_2}|<4.6\cdot 10^{-14}\gev \qquad
(95\%\text{ C.L.}).
\ee
A detailed analysis of the implications of this bound on the mass spectrum 
of mirror fermions has been presented in \cite{Hubisz}. We do not want to 
repeat this analysis here as we basically agree with the results of these 
authors in this case. In all our numerical results the bound in 
(\ref{Dbound}) has been taken into account.

\newsection{Summary and Outlook}
\label{sec:summary}

In this paper we have
calculated a number of observables related to particle-antiparticle mixing in 
the Littlest Higgs model (LHT) with T-parity. The first analysis 
of particle-antiparticle mixing in this model has been presented by 
Hubisz et al.~\cite{Hubisz}. We confirm  the effective 
Hamiltonian for $\Delta F=2$ transitions found by these authors 
but our phenomenological analysis differs from theirs in various aspects. 
While Hubisz et al.~studied only $\Delta M_q$ mass differences and 
$\varepsilon_K$ with the goal to constrain the mass spectrum and weak 
mixing matrix of mirror fermions, our main goal was to include in the 
analysis also most interesting CP-violating observables in $B_d$ and 
$B_s$ decays and to use  the new flavour and CP-violating interactions 
present in the LHT model to remove possible discrepancies between the SM
and existing data. Moreover, we have calculated in the LHT model for the first time 
the branching ratios for the $B\to X_{s,d}\gamma$ decays and the related CP asymmetries.

The main messages of our paper are as follows:
\begin{itemize}
\item
The LHT model can be made consistent with all FCNC processes considered 
in the present paper for masses of mirror fermions and new weak
gauge bosons in the reach of LHC, provided the weak mixing matrix 
$V_{Hd}$ exhibits a hierarchical structure and the mass spectrum of 
mirror fermions is quasi-degenerate.
\item
We emphasize, however, that the structure of the mixing matrix 
$V_{Hd}$ can differ 
significantly from the known structure of the CKM matrix so that 
interesting departures from MFV correlations between various 
processes are possible. Basically all MFV correlations between $K$, $B^0_d$ 
and $B^0_s$ meson systems can be modified, while being still consistent 
with the existing data, even if these modifications 
amount to at most $30\%$ in the case of the CP-conserving observables considered here.
\item
The above size of still possible deviations from the SM implies that 
the mass differences $\Delta M_q$ and $\varepsilon_K$ considered in 
\cite{Hubisz} are not the appropriate observables to identify possible 
signals from mirror fermions, heavy gauge bosons and $T_+$,
 as the non-perturbative uncertainties in these observables are comparable
to the new effects themselves. 
A good example are the results in (\ref{DMsSM}).
Certainly, $\Delta M_q$ and $\varepsilon_K$ 
can serve as first tests of the viability of the model but to constrain and 
test 
the model in detail, significantly cleaner, from the theoretical point 
of view, observables have to be considered. These are in particular the 
mixing induced CP asymmetries $S_{\psi K_S}$ and $S_{\psi \phi}$ but 
also $\Delta M_d/\Delta M_s$, $A_{\rm SL}^q$ and $\Delta \Gamma_q$. 
This also applies to $Br(B\to X_{s,d}\gamma)$, the related CP asymmetries and 
a number of rare decay branching ratios with the latter considered in 
a separate paper \cite{BBPRTUW}.
\item
We find that the T-even sector of the LHT model, that represents this 
model in FCNC processes in the limit of exactly degenerate mirror fermions
is not favoured by the data as independently of the parameters of this sector 
$\Delta M_s > (\Delta M_s)_{\rm SM}$ and the possible discrepancy between 
the value of the CP asymmetry $S_{\psi K_S}$ and large 
values of $|V_{ub}|$ cannot be removed.
\item
Using the full structure of new flavour and CP-violating interactions 
encoded in $V_{Hd}\not=V_{\rm CKM}$, we identify regions in the parameter
space of the LHT model in which possible problems of the SM can be 
cured, large CP-violating effects in the $B^0_s$ system are predicted
 and the mass difference 
$\Delta M_s$ is found to be smaller than $(\Delta M_s)_{\rm SM}$ as
suggested by the recent result of  the CDF collaboration. 
\item
In particular we  identify a scenario  in which 
significant enhancements of the CP asymmetries 
$S_{\psi \phi}$ and $A_{\rm SL}^q$ relative to the  SM
are possible, while satisfying all existing 
constraints, in particular from the $B\to X_s\gamma$ decay and 
$A_{\rm CP}(B\to X_s\gamma)$ that are presented in the LHT model here
for the first time. 
In this scenario the weak mixing matrix of mirror fermions turns out to 
have a hierarchical structure that differs by much from the CKM one.
\item
In another scenario the second, non-SM, value for the angle 
$\gamma=-(109\pm16)^\circ$ from tree level decays can be
made consistent with all existing data with the help of mirror fermions.
\item
We have found
a number of correlations between the observables in question
and studied the
implications of our results for the mass spectrum and the weak mixing 
matrix of mirror fermions. 
\item
The effects from mirror fermions in  the
 $B\to X_s\gamma$ decay turn out to be smaller than in the $\Delta B=2$ 
 transitions, which should be welcomed as the SM is here in a rather 
good shape. Typically the new physics effects are below $4\%$.
\item
We also find that the new physics effects in 
$A_{\rm CP}(B\to X_{s,d}\gamma)$ are very small but their measurements
could in principle help to rule out the $\gamma=-109^\circ$ solution
from tree level decays, as $A_{\rm CP}(B\to X_{s,d}\gamma)$ reverses its sign.
A similar comment applies to $A_\text{SL}^d$.
\end{itemize}

\subsection*{Acknowledgments}

We would like to thank Marcella Bona, Gino Isidori and Luca Silvestrini for
useful discussions and Stefan Recksiegel for a careful reading of the paper.
A.J.B. and A.W. would also like to thank Luca Silvestrini and INFN for the 
hospitality at the University of Rome ``La Sapienza'', where the final steps of this paper have been made.  This research was partially supported by the German `Bundesministerium f\"ur 
Bildung und Forschung' under contract 05HT4WOA/3. 

\newpage
\begin{appendix}

\newsection{Non-leading Contributions of $\bm{T_-}$ and $\bm{\Phi}$}
\label{sec:appA}

Here we want to demonstrate explicitly that the T-odd heavy
$T_-$ does not contribute to any of the processes 
we study and that the heavy scalar triplet $\Phi$ does not contribute 
at $\ord(v^2/f^2)$. 
The reasons are as follows:
\begin{itemize}
\item Omitting the first two quark generations, 
the masses for $t$, $T_+$ and $T_-$ are generated through the
  following Yukawa interaction \cite{tparity,mirror}:
\bea\nn
\mathcal{L}_\text{top}&=&-\frac{1}{2\sqrt{2}}\lambda_1 f
  \epsilon_{ijk}\epsilon_{xy}\left[(\bar Q_1)_i(\Sigma)_{jx}(\Sigma)_{ky} - (\bar Q_2 \Sigma_0)_i (\tilde
    \Sigma)_{jx}(\tilde \Sigma)_{ky}\right] t_R \\
&&-\; \lambda_2 f (\bar
  t'_1 t'_{1R}+ \bar t'_2 t'_{2R}) +h.c.\label{upYuk}\,.
\eea
This leads to a mixing between the weak eigenstates of $t$
and $T_+$, and therefore, couplings of the form $\bar T_+ W_L^+
d^i$ exist. They are suppressed by $v/f$, as the mixing appears
only at this order.
\item However, $u^3_H$, as all other mirror fermions, gets its mass
  from the Dirac mass term (omitting again the first two generations) \cite{mirror}
\be
\mathcal{L}_\text{Dirac}=-\kappa
f\left(\bar\Psi_{2}\xi\Psi_R+\bar\Psi_{1}\Sigma_0\Omega\xi^\dagger\Omega\Psi_R\right)+h.c.
\ee
so that there is no tree level mixing of $T_-$ with $u^3_H$ (and the other
mirror quarks). Therefore, $T_-$ stays singlet under
$SU(2)_1\times SU(2)_2$ and does not couple to ordinary down-type
quarks. Thus $T_-$ does not contribute neither to $\Delta B=2$ and
$\Delta S=2$ processes nor to $B\to X_s \gamma$.
\item The case of $D^0-\bar D^0$ mixing is slightly more
  involved. Here, $T_-$ could contribute via interactions $\bar q A_H
  T_-$ and  $\bar q Z_H  T_-$ ($q=u,c$): $T_-$ couples to the weak
  eigenstate of $T_+$ through the interaction with $B_H$. As $T_+$
  gets its mass from the up-type Yukawa term \eqref{upYuk}, which
  also generates the masses of the three up-type quarks, it can in principle
  mix with all three of them, as pointed out in \cite{Lee:2004me}. However, as found there, this mixing is highly constrained for the first two
  generations, so we can safely neglect it. In this approximation,
  there are thus no couplings of the form $\bar q A_H
  T_-$ and  $\bar q Z_H  T_-$ ($q=u,c$).
\item In summary we find that $T_-$ has a sizable flavour changing
  coupling only to $t$, thus confirming the corresponding statement made in \cite{Hubisz}.
\end{itemize}

For completeness, we also have to consider the contributions of
the scalar triplet $\Phi$ to the processes analyzed in the present
paper. The relevant diagrams can be obtained by simply replacing
$W_H^\pm$ by $\phi^\pm$ and $A_H,\;Z_H$ by $\phi^0,\;\phi^P$ in
the diagrams shown in Figs.~\ref{fig:mixodd} and \ref{fig:bsgodd}.
However, all couplings of $\Phi$ to fermions turn out to be
$\ord(v/f)$, so that the effect of those diagrams is of higher
order in $v/f$ than the one resulting from diagrams with gauge
boson exchanges. Therefore the scalar triplet $\Phi$ does not
contribute at $\ord(v^2/f^2)$ to the processes in question in the 
LHT model.

\newsection{Relevant Functions}
\label{sec:appB}

In this Appendix we list the functions that entered the present
study of $\Delta F=2$ and $B \rightarrow X_s \gamma$ processes.
Both the SM contributions and the new physics contributions coming
from the T-even and T-odd sectors are collected. The variables are
defined as follows: \bea
&x_q=\dfrac{m_q^2}{M_{W_L}^2}\,,\qquad x_T=\dfrac{m_{T_+}^2}{M_{W_L}^2}\,\qquad(q=c,t)\,,&\nn\\
&z_i=\dfrac{m_{Hi}^2}{M_{W_H}^2}\,,\qquad
z'_i=\dfrac{m_{Hi}^2}{M_{A_H}^2}=z_i\,a \qquad \text{with}\quad
a=\dfrac{5}{\tan^2 \theta_W}\,\qquad (i=1,2,3)\,.& \eea

\subsection{\boldmath Functions entering $\Delta F=2$ Processes\unboldmath}

\begin{eqnarray}
S_0(x_t)&=&\frac{x_t\,( 4 - 11\,x_t + x_t^2) }
   {4\,{( -1 + x_t ) }^2} +
  \frac{3\,x_t^3\,\log x_t}{2\,{( -1 + x_t ) }^3}\\
\nonumber\\
S_0(x_c, x_t)&=&\frac{-3 x_t x_c}
   {4 ( -1 + x_t)( -1 + x_c ) } -
  \frac{x_t( 4 - 8 x_t + x_t^2 ) x_c \log x_t}
   {4 {( -1 + x_t) }^2
     ( -x_t + x_c ) }\nonumber\\
&& +
  \frac{x_t x_c ( 4-8 x_c +x_c ^2 ) \log x_c}
   {4 {( -1 + x_c ) }^2
     ( -x_t + x_c ) }\\
\nonumber\\
P_{1}(x_{t}, x_{T}) & = & \frac{x_{t}(-4 + 11 x_{t} - x_{t}^{2} + x_{T} - 8 x_{t} x_{T} + x_{t}^{2} x_{T})}{4(-1 + x_{t})^{2}(-1 + x_{T})} + \frac{x_{t} x_{T}(4 - 8 x_{T} + x_{T}^{2}) \log{x_{T}}}{4(x_{t} - x_{T}) (-1 + x_{T})^{2}}
\nonumber \\
&& -  \frac{x_{t}(-6 x_{t}^{3} - 4 x_{T} + 12 x_{t} x_{T} - 3
  x_{t}^{2} x_{T} + x_{t}^{3} x_{T}) \log{x_{t}}}{4(-1 +
  x_{t})^{3}(x_{t} - x_{T})}\\
\nonumber\\
P_{2}(x_{c}, x_{t}, x_{T}) & = & \frac{3 (x_{t} x_{c} - x_{T}
  x_{c})}{4(-1 + x_{t})(-1 + x_{T}) (-1 + x_{c})} + \frac{(4 x_{t}
  x_{c} - 8 x_{t}^{2} x_{c} + x_{t}^{3} x_{c}) \log{x_{t}}}{4(-1 +
  x_{t})^{2}(x_{t} - x_{c})} \nonumber \\
&& +  \frac{(4 x_{t} x_{c}^{2} - 4 x_{T} x_{c}^{2} - 8  x_{t} x_{c}^{3} + 8
   x_{T} x_{c}^{3} + x_{t} x_{c}^{4} - x_{T} x_{c}^{4}) \log{x_{c}}}{4(x_{t} -
   x_{c})(x_{T} - x_{c}) (-1 + x_{c})^{2}}\nonumber \\
&& -  \frac{(4 x_{T} x_{c} - 8 x_{T}^{2} x_{c} + x_{T}^{3} x_{c})
  \log{x_{T}}}{4(-1 + x_{T})^{2}(x_{T} - x_{c})}
\end{eqnarray}
\begin{eqnarray}
F(z_i,z_j;W_H)&=& \frac{1}{(1-z_i)(1-z_j)} \left(1-\frac{7}{4} z_i
z_j\right) +\frac{z_i^2 \log z_i}{(z_i - z_j) (1-z_i)^2} \left( 1- 2
z_j + \frac{z_i z_j}{4} \right)\nn\\
&& -\frac{z_j^2 \log z_j}{(z_i - z_j) (1-z_j)^2} \left( 1- 2
z_i + \frac{z_i z_j}{4} \right) \\\nn\\
G(z_i,z_j;Z_H) &=& -\frac{3}{4} \left[\frac{1}{(1-z_i)(1-z_j)}+\frac{z_i^2 \log
    z_i}{(z_i - z_j) (1-z_i)^2} - \frac{z_j^2 \log z_j}{(z_i - z_j)
    (1-z_j)^2} \right]\\\nn\\
A_1 (z_i, z_j ; Z_H ) &=& -\frac{3}{100 a} \left[ \frac{1}{(1- z'_i)(1-z'_j)} + \frac{z'_i z_i \log
    z'_i}{(z_i-z_j) (1-z'_i)^2}\right. \nn\\
&&\left.  - \frac{z'_j z_j \log
    z'_j}{(z_i-z_j) (1-z'_j)^2} \right] \\\nn\\
A_2 (z_i, z_j ; Z_H ) &=& -\frac{3}{10} \left[ \frac{\log a}{(a-1) (1- z'_i)(1-z'_j)} + \frac{z_i^2 \log
    z_i}{(z_i-z_j) (1-z_i) (1- z'_i)}\right. \nn\\
&&\left.  - \frac{z_j^2 \log
    z_j}{(z_i-z_j) (1-z_j)(1-z'_j)} \right],
\end{eqnarray}

\subsection{\boldmath Functions entering $B\to X_s\gamma$\unboldmath}

\bea D'_0(y)&=&-\dfrac{(3y^3-2y^2)}{2(y-1)^4}\log y +
\dfrac{(8y^3+5y^2-7y)}{12(y-1)^3}\\
E'_0(y)&=&\dfrac{3y^2}{2(y-1)^4}\log y +
\dfrac{(y^3-5y^2-2y)}{4(y-1)^3}\qquad(y=x_t,x_T,z_i,z'_i)
\eea

\end{appendix}

\end{document}